\documentclass[aps,prd,preprint,superscriptaddress,showpacs,floatfix,nobibnotes,nofootinbib]{revtex4}
\usepackage{mathrsfs}

\usepackage{bm}
\usepackage{graphicx}
\usepackage{longtable}
\usepackage{amsmath}
\usepackage{amssymb}

\begin{document}

\title{Calculation of mass and width of unstable molecular state using the developed Bethe-Salpeter theory}

\author{Xiaozhao Chen}\email{chen_xzhao@sina.com}
\email[corresponding author]{} \affiliation{Department of Fundamental Courses, Shandong University of Science and Technology, Taian, 271019, China}

\author{Xiaofu L\"{u}}
\affiliation{Department of Physics, Sichuan University, Chengdu, 610064, China}
\affiliation{Institute of Theoretical Physics, The Chinese Academy of Sciences, Beijing 100080, China}
\affiliation{CCAST (World Laboratory), P.O. Box 8730, Beijing  100080, China}

\author{Xiurong Guo}
 \affiliation{Department of Fundamental Courses, Shandong University of Science and Technology, Taian, 271019, China}

\author{Zonghua Shi}
 \affiliation{Department of Fundamental Courses, Shandong University of Science and Technology, Taian, 271019, China}

\author{Qingbiao Wang}
 \affiliation{Department of Fundamental Courses, Shandong University of Science and Technology, Taian, 271019, China}

\date{\today}

\begin{abstract}
Applying the developed Bethe-Salpeter theory for dealing with resonance, we investigate the time evolution of molecular state composed of two vector mesons as determined by the total Hamiltonian. Then exotic meson resonance $\chi_{c0}(3915)$ is considered as a mixed state of two unstable molecular states $D^{*0}\bar{D}^{*0}$ and $D^{*+}D^{*-}$, and the mass and width for physical resonance $\chi_{c0}(3915)$ are calculated in the framework of relativistic quantum field theory. In this actual calculation, we minutely show how to obtain the correction for energy level of resonance and to exhibit the key features of dispersion relation in an extended Feynman diagram. The numerical results are consistent with the experimental values.
\end{abstract}

\pacs{12.40.Yx, 14.40.Rt, 12.39.Ki}


\maketitle

\newpage

\parindent=20pt

\section{introduction}
Hadronic molecule structure has been proposed to interpret the internal structure of exotic meson resonance for many years \cite{ms:Swanson,ms:Torn}. In the previous works \cite{ms:Swanson,ms:Torn,liu,mypaper4,Msigma2}, molecular states were considered as meson-meson bound states and homogeneous Bethe-Salpeter (BS) equation was frequently used to investigate molecular states. Solving homogeneous BS equations for meson-meson bound states, the authors of these works obtained the masses and BS wave functions. The mass of meson-meson bound state was regarded as mass of exotic meson resonance. However, all decay channels of resonance should contribute to its physical mass and the correction for energy level of molecular state due to decay channels has seldom been considered \cite{ms:Swanson,ms:Torn,liu,mypaper4,Msigma2,mypaper6,mypaper7}. Fortunately, recent fundamental research \cite{mypaper8} has noticed that hadron resonance should be regarded as an unstable two-body system, and developed BS theory for dealing with the dynamics of coupled channels in the framework of relativistic quantum field theory. Though Ref. \cite{mypaper8} illuminated the physical meaning of the developed Bethe-Salpeter theory for dealing with resonance, many details in computational process have not been presented. In this paper, we will comprehensively and systematically show the theoretical approach about unstable molecular state composed of two heavy vector mesons, and this approach is applied to investigate exotic meson resonance $\chi_{c0}(3915)$ \cite{mutlistate}, once named $\emph{X}(3915)$, which is considered as a mixed state of two unstable molecular states $D^{*0}\bar{D}^{*0}$ and $D^{*+}D^{*-}$.

Since resonance is an unstable state which decays spontaneously into other particles, the molecular state composed of two heavy vector mesons should not be a stationary vector-vector bound state. To investigate this unstable two-body system, we suppose that at some given time this unstable state has been prepared to decay and then study the time evolution of this system as determined by the total Hamiltonian. This prepared state can be described by the ground-state BS wave function for vector-vector bound state at the times $t_1=0$ and $t_2=0$. In our previous works \cite{mypaper4,mypaper7}, the most general form of BS wave functions for the bound states created by two vector fields with arbitrary spin and definite parity has been given. According to the effective theory at low energy QCD, we have investigated the light meson interaction with quarks in heavy vector mesons and obtained the interaction kernel between two heavy vector mesons derived from one light meson ($\sigma$, $\omega$, $\rho$, $\phi$) exchange \cite{mypaper3,mypaper4}. Solving BS equation with this interaction kernel, we have obtained the mass and BS wave function for bound state composed of two vector mesons \cite{mypaper4,mypaper5}. After providing the description for the prepared state, we can study the time evolution of BS wave function and obtain the pole corresponding to resonance through the scattering matrix element.

The crucial point of our resonance theory is that the scattering matrix element between bound states is calculated in the framework of relativistic quantum field theory. According to dispersion relation, the total matrix element between a final state and an initial bound state should be calculated with respect to arbitrary value of the final state energy \cite{mypaper8}. It is necessary to note that the total energy of the final state extends over the real interval while the initial state energy is specified. For the initial bound state composed of two heavy vector mesons, we have given the generalized Bethe-Salpeter (GBS) amplitude for four-quark state describing this meson-meson structure \cite{mypaper6,mypaper7}, which should be specified. Because the value of the final state energy is an arbitrary real number over the real interval, we may obtain several closed channels derived from the interaction Lagrangian and all open and closed channels should contribute to the mass of physical resonance. For exotic resonance $\chi_{c0}(3915)$, we consider three open decay channels $J/\psi\omega$, $D^+D^-$, $D^0\bar D^0$ and one closed channel $D^{*}\bar D^{*}$ from the effective interaction Lagrangian at low energy QCD. Mandelstam's approach is applied to calculate the matrix element between bound states with respect to arbitrary value of the final state energy, which are exhibited by extended Feynman diagrams. Finally, we obtain the correction for energy level of resonance $\chi_{c0}(3915)$ and the physical mass is used to calculate the decay width of physical resonance $\chi_{c0}(3915)$.

The structure of this article is as follows. In Sec. \ref{sec:GBSwfmm} we give the revised general form of GBS wave functions for meson-meson bound states as four-quark states. The mass and GBS wave function for the mixed state of two bound states $D^{*0}\bar{D}^{*0}$ and $D^{*+}D^{*-}$ is obtained in instantaneous approximation. Section \ref{sec:Sme} gives the traditional technique to calculate the matrix element with mass of meson-meson bound state, which is applied to investigate the decay modes $\chi_{c0}(3915)\rightarrow J/\psi\omega$, $\chi_{c0}(3915)\rightarrow D^+D^-$ and $\chi_{c0}(3915)\rightarrow D^0\bar D^0$. Section \ref{sec:tetotalH} gives the developed Bethe-Salpeter theory. In Sec. \ref{sec:ac} we emphatically introduce the matrix element between bound states with respect to arbitrary value of the final state energy. Three open decay channels $J/\psi\omega$, $D^+D^-$, $D^0\bar D^0$ and one closed channel $D^{*}\bar D^{*}$ are considered. In Sec. \ref{sec:mw} we obtain the physical mass and width for unstable molecular state. Our numerical results are presented in Sec. \ref{sec:nr} and we make some concluding remarks in Sec. \ref{sec:concl}.

\section{GBS wave function of meson-meson bound state as a four-quark state}\label{sec:GBSwfmm}
In this paper, we investigate the light meson interaction with the light quarks in heavy mesons. As in effective theory at low energy QCD, the interaction Lagrangian for the coupling of light quark fields to light meson fields is \cite{mypaper6}
\begin{equation}\label{Lag}
\begin{split}
&\mathscr{L}^{\text{eff}}_I=ig_0\left(\begin{array}{ccc} \bar{u}&\bar{d}&\bar{s}
\end{array}\right)\gamma_5\left(\begin{array}{ccc}
\pi^0+\frac{1}{\sqrt{3}}\eta &\sqrt{2}\pi^+&\sqrt{2}K^+\\\sqrt{2}\pi^-&-\pi^0+\frac{1}{\sqrt{3}}\eta &\sqrt{2}K^0\\\sqrt{2}K^-&\sqrt{2}\bar{K}^0&-\frac{2}{\sqrt{3}}\eta
\end{array}\right)\left(\begin{array}{c} u\\d\\s
\end{array}\right)\\
&+ig'_0\left(\begin{array}{ccc} \bar{u}&\bar{d}&\bar{s}
\end{array}\right)\gamma_\mu\left(\begin{array}{ccc}
\rho^0+\omega&\sqrt{2}\rho^+&\sqrt{2}K^{*+}\\\sqrt{2}\rho^-&-\rho^0+\omega&\sqrt{2}K^{*0}\\\sqrt{2}K^{*-}&\sqrt{2}\bar{K}^{*0}&\sqrt{2}\phi
\end{array}\right)_\mu\left(\begin{array}{c} u\\d\\s
\end{array}\right)+g_\sigma\left(\begin{array}{cc} \bar{u}&\bar{d}
\end{array}\right)\left(\begin{array}{c} u\\d
\end{array}\right)\sigma.
\end{split}
\end{equation}
From this effective interaction Lagrangian at low energy QCD, we have to consider that the heavy meson is a bound state composed of a quark and an antiquark and investigate the interaction of light meson with quarks in heavy meson. The quark current $J_\mu$ coupling with light vector meson and the quark scalar density $J$ coupling with $\sigma$ meson can be obtained. In this section, our attention is only focused on the bound state composed of two vector mesons and some errors in previous works are revised.

\subsection{BS wave function for bound state composed of two vector mesons}\label{sec:BS}
If a bound state with spin $j$ and parity $\eta_{P}$ is created by two Heisenberg vector fields with masses $M_1$ and $M_2$, respectively, its BS wave function is defined as
\begin{equation}\label{BSwfdx}
\chi_{P(\lambda\tau)}^j(x_1',x_2')=\langle0|TA_\lambda(x_1')A^\dagger_\tau(x_2')|P,j\rangle=\frac{1}{(2\pi)^{3/2}}\frac{1}{\sqrt{2E(P)}}e^{iP\cdot X}\chi_{P(\lambda\tau)}^j(X'),
\end{equation}
where $P$ is the momentum of the bound state, $E(p)=\sqrt{\mathbf{p}^2+m^2}$, $x_1'=(\mathbf{x}_1',it_1)$, $x_2'=(\mathbf{x}_2',it_2)$, $X=\eta_1x_1'+\eta_2x_2'$, $X'=x_1'-x_2'$ and $\eta_{1,2}=M_{1,2}/(M_1+M_2)$. Making the Fourier transformation, we obtain the BS wave function in the momentum representation
\begin{equation}\label{BSwfdp}
\begin{split}
\chi^j_P(p_1',p_2')_{\lambda\tau}=\frac{1}{(2\pi)^{3/2}}\frac{1}{\sqrt{2E(P)}}(2\pi)^4\delta^{(4)}(P-p_1'+p_2')\chi^j_{\lambda\tau}(P,p),
\end{split}
\end{equation}
where $p$ is the relative momentum of two vector fields and we have $P=p_1'-p_2'$, $p=\eta_2p_1'+\eta_1p_2'$, $p_1'$ and $p_2'$ are the momenta carried by two vector fields, respectively. The polarization tensor of the bound state $\eta_{\mu_1\mu_2\cdots\mu_j}$ can be separated,
\begin{equation}\label{arbspin}
\chi_{\lambda\tau}^j(P,p)=\eta_{\mu_1\mu_2\cdots\mu_j}\chi_{\mu_1\mu_2\cdots\mu_j\lambda\tau}(P,p),
\end{equation}
where the subscripts $\lambda$ and $\tau$ are derived from these two vector fields. The polarization tensor $\eta_{\mu_1\mu_2\cdots\mu_j}$ describes the spin of the bound state, which is totally symmetric, transverse and traceless:
\begin{equation}\label{polar}
\eta_{\mu_1\mu_2\cdots}=\eta_{\mu_2\mu_1\cdots},\quad P_{\mu_1}\eta_{\mu_1\mu_2\cdots}=0,\quad \eta_{\mu_1\mu_2\cdots}^{\mu_1}=0.
\end{equation}
From Lorentz covariance, we have
\begin{equation}\label{scalarfun}
\begin{split}
\chi_{\mu_1\cdots\mu_j\lambda\tau}=&p_{\mu_1}\cdots p_{\mu_j}[g_{\lambda\tau}f_1+(P_\lambda p_\tau+P_\tau p_\lambda)f_2 +(P_\lambda p_\tau-P_\tau p_\lambda)f_3+P_\lambda P_\tau f_4+p_\lambda p_\tau f_5]\\
&+(p_{\{\mu_2}\cdots p_{\mu_j}g_{\mu_1\}\lambda}p_\tau+p_{\{\mu_2}\cdots p_{\mu_j}g_{\mu_1\}\tau} p_\lambda)f_6\\
&+(p_{\{\mu_2}\cdots p_{\mu_j}g_{\mu_1\}\lambda}p_\tau-p_{\{\mu_2}\cdots p_{\mu_j}g_{\mu_1\}\tau} p_\lambda)f_7\\
&+(p_{\{\mu_2}\cdots p_{\mu_j}g_{\mu_1\}\lambda}P_\tau+p_{\{\mu_2}\cdots p_{\mu_j}g_{\mu_1\}\tau} P_\lambda)f_8\\
&+(p_{\{\mu_2}\cdots p_{\mu_j}g_{\mu_1\}\lambda}P_\tau-p_{\{\mu_2}\cdots p_{\mu_j}g_{\mu_1\}\tau} P_\lambda)f_9\\
&+p_{\mu_1}\cdots p_{\mu_j}\epsilon_{\lambda\tau\xi\zeta}p_\xi P_\zeta f_{10}+p_{\{\mu_2}\cdots p_{\mu_j}\epsilon_{\mu_1\}\lambda\tau\xi}p_\xi f_{11}+p_{\{\mu_2}\cdots p_{\mu_j}\epsilon_{\mu_1\}\lambda\tau\xi}P_\xi f_{12}\\
&+(p_{\{\mu_2}\cdots p_{\mu_j}\epsilon_{\mu_1\}\lambda\xi\zeta}p_\xi P_\zeta p_\tau+p_{\{\mu_2}\cdots p_{\mu_j}\epsilon_{\mu_1\}\tau\xi\zeta}p_\xi P_\zeta p_\lambda)f_{13}\\
&+(p_{\{\mu_2}\cdots p_{\mu_j}\epsilon_{\mu_1\}\lambda\xi\zeta}p_\xi P_\zeta p_\tau-p_{\{\mu_2}\cdots p_{\mu_j}\epsilon_{\mu_1\}\tau\xi\zeta}p_\xi P_\zeta p_\lambda)f_{14}\\
&+(p_{\{\mu_2}\cdots p_{\mu_j}\epsilon_{\mu_1\}\lambda\xi\zeta}p_\xi P_\zeta P_\tau+p_{\{\mu_2}\cdots p_{\mu_j}\epsilon_{\mu_1\}\tau\xi\zeta}p_\xi P_\zeta P_\lambda)f_{15}\\
&+(p_{\{\mu_2}\cdots p_{\mu_j}\epsilon_{\mu_1\}\lambda\xi\zeta}p_\xi P_\zeta P_\tau-p_{\{\mu_2}\cdots p_{\mu_j}\epsilon_{\mu_1\}\tau\xi\zeta}p_\xi P_\zeta P_\lambda)f_{16}\\
&+p_{\{\mu_3}\cdots p_{\mu_j}g_{\mu_1\lambda}g_{\mu_2\}\tau}f_{17}+p_{\{\mu_3}\cdots p_{\mu_j}\epsilon_{\mu_{1}\lambda\xi\zeta}p_\xi P_\zeta\epsilon_{\mu_{2}\}\tau\xi'\zeta'}p_{\xi'} P_{\zeta'} f_{18}\\
&+(p_{\{\mu_3}\cdots p_{\mu_j}g_{\mu_1\lambda}\epsilon_{\mu_2\}\tau\xi\zeta}p_\xi P_\zeta+p_{\{\mu_3}\cdots p_{\mu_j}g_{\mu_1\tau}\epsilon_{\mu_2\}\lambda\xi\zeta}p_\xi P_\zeta)f_{19}\\
&+(p_{\{\mu_3}\cdots p_{\mu_j}g_{\mu_1\lambda}\epsilon_{\mu_2\}\tau\xi\zeta}p_\xi P_\zeta-p_{\{\mu_3}\cdots p_{\mu_j}g_{\mu_1\tau}\epsilon_{\mu_2\}\lambda\xi\zeta}p_\xi P_\zeta)f_{20},
\end{split}
\end{equation}
where $\{\mu_1,\cdots,\mu_j\}$ represents symmetrization of the indices $\mu_1,\cdots,\mu_j$. In fact, the relative momenta $p_{\mu_1},\cdots,p_{\mu_j},p_{\lambda},p_{\tau}$ represent the orbital angular momenta. There should be 20 scalar functions $f_i(P\cdot p,p^2)(i=1,\cdots,20)$ in Eq. (\ref{scalarfun}). In Ref. \cite{mypaper7}, three tensor structures are omitted. In this paper, these missing terms are added as the last three terms in Eq. (\ref{scalarfun}). Using the transversality condition \cite{mypaper3,mypaper4}
\begin{equation}\label{proofvec}
p'_{1\lambda}\chi^j_{\lambda\tau}(P,p)=p'_{2\tau}\chi^j_{\lambda\tau}(P,p)=0
\end{equation}
and considering the properties of BS wave function under space reflection, we obtain the revised general form of BS wave functions for the bound states created by two massive vector fields with arbitrary spin and definite parity (see details in \cite{mypaper4}),
for $\eta_{P}=(-1)^j$,
\begin{equation}\label{jp}
\begin{split}
\chi_{\lambda\tau}^{j}(P,p)=&\frac{1}{\mathcal{N}^j}\eta_{\mu_1\cdots\mu_j}[p_{\mu_1}\cdots p_{\mu_j}(\mathcal{T}^1_{\lambda\tau}\Phi_1+\mathcal{T}^2_{\lambda\tau}\Phi_2)+\mathcal{T}^3_{\mu_1\cdots\mu_j\lambda\tau}\Phi_3+\mathcal{T}^4_{\mu_1\cdots\mu_j\lambda\tau}\Phi_4\\
&+\mathcal{T}^5_{\mu_1\cdots\mu_j\lambda\tau}\Phi_5+\mathcal{T}^6_{\mu_1\cdots\mu_j\lambda\tau}\Phi_6],
\end{split}
\end{equation}
for $\eta_{P}=(-1)^{j+1}$,
\begin{equation}\label{jm}
\begin{split}
\chi_{\lambda\tau}^{j}(P,p)=&\frac{1}{\mathcal{N}^j}\eta_{\mu_1\cdots\mu_j}(p_{\mu_1}\cdots p_{\mu_j}\epsilon_{\lambda\tau\xi\zeta}p_\xi P_\zeta\Phi'_1+\mathcal{T}^7_{\mu_1\cdots\mu_j\lambda\tau}\Phi'_2+\mathcal{T}^8_{\mu_1\cdots\mu_j\lambda\tau}\Phi'_3+\mathcal{T}^{9}_{\mu_1\cdots\mu_j\lambda\tau}\Phi'_4\\
&+\mathcal{T}^{10}_{\mu_1\cdots\mu_j\lambda\tau}\Phi'_5+\mathcal{T}^{11}_{\mu_1\cdots\mu_j\lambda\tau}\Phi'_6+\mathcal{T}^{12}_{\mu_1\cdots\mu_j\lambda\tau}\Phi'_7),
\end{split}
\end{equation}
where $\mathcal{N}^j$ is normalization, the independent tensor structures $\mathcal{T}^i_{\lambda\tau}$ are given in Appendix \ref{app1}, $\Phi_i(P\cdot p,p^2)$ and $\Phi'_i(P\cdot p,p^2)$ are independent scalar functions. Scalar functions $f_i$ in Eq. (\ref{scalarfun}) are the linear combinations of $\Phi_i$ and $\Phi'_i$.

\subsection{Kernel between two heavy vector mesons}\label{sec:kernel}
In this paper, we assume that the isoscalar $\chi_{c0}(3915)$ is a mixed state of two unstable molecular states $D^{*0}\bar{D}^{*0}$ and $D^{*+}D^{*-}$ with spin-parity quantum numbers $0^+$. In this section, we only investigate the mixed state of two stable bound states $D^{*0}\bar{D}^{*0}$ and $D^{*+}D^{*-}$, and BS wave function for this system is a linear combination of two components as
\begin{equation}\label{mmbounstate}
\begin{split}
\chi^{D^{*}\bar D^{*},j}_{\lambda\tau}(P,p)=&\frac{1}{\sqrt{2}}\chi^{D^{*0}\bar D^{*0},j}_{\lambda\tau}(P,p)+\frac{1}{\sqrt{2}}\chi^{D^{*+}D^{*-},j}_{\lambda\tau}(P,p),
\end{split}
\end{equation}
where
\begin{equation}\label{isoscalarBSwf2}
\begin{split}
\chi^{D^{*0}\bar D^{*0},j}_{\lambda\tau}(P,p)=&\chi^{j}_{\lambda\tau}(P,p)\bigg(-\bigg|\frac{1}{2},-\frac{1}{2}\bigg\rangle\bigg)^{D^{*0}}\otimes\bigg|\frac{1}{2},\frac{1}{2}\bigg\rangle^{\bar D^{*0}},\\
\chi^{D^{*+}D^{*-},j}_{\lambda\tau}(P,p)=&\chi^{j}_{\lambda\tau}(P,p)\bigg(-\bigg|\frac{1}{2},\frac{1}{2}\bigg\rangle\bigg)^{D^{*+}}\otimes\bigg|\frac{1}{2},-\frac{1}{2}\bigg\rangle^{D^{*-}},
\end{split}
\end{equation}
and $P$ becomes the total momentum for the mixed state of two meson-meson bound states, $\chi^{j}_{\lambda\tau}(P,p)$ is the component wave function in the momentum representation; $(-|\frac{1}{2},-\frac{1}{2}\rangle)\otimes|\frac{1}{2},\frac{1}{2}\rangle$ and $(-|\frac{1}{2},\frac{1}{2}\rangle)\otimes|\frac{1}{2},-\frac{1}{2}\rangle$ are the isospin wave functions of pure bound states $D^{*0}\bar D^{*0}$ and $D^{*+}D^{*-}$, respectively; $\chi^{D^{*0}\bar D^{*0},j}_{\lambda\tau}$ and $\chi^{D^{*+}D^{*-},j}_{\lambda\tau}$ represent BS wave functions for the bound states of two vector mesons, which are the eigenstates of Hamiltonian without considering the coupled-channel terms. These eigenstates have the same quantum numbers. The error in Ref. \cite{mypaper7} has been revised. As usual the momentum for the mixed state of two bound states is set as $P=(0,0,0,iM_0)$ in the rest frame.

Let $D^*_l$ denote one of $D^{*0}$ and $D^{*+}$, and $l=u,d$ represents the $u$ or $d$ antiquark in heavy vector meson $D^{*0}$ or $D^{*+}$, respectively; $\bar{D}_l^{*}$ denotes the antiparticle of $D_l^{*}$. From Eq. (\ref{jp}), we can obtain BS wave function describing pure bound state $D^{*}_l\bar D^{*}_l$
\begin{equation}\label{bswf0+}
\begin{split}
\chi_{\lambda\tau}^{0^+}(P^{D\bar D},p)=\frac{1}{\mathcal{N}_{D\bar D}^{0^+}}[\mathcal{T}^1_{\lambda\tau}\mathcal{F}_1(P^{D\bar D}\cdot p,p^2)+\mathcal{T}^2_{\lambda\tau}\mathcal{F}_2(P^{D\bar D}\cdot p,p^2)].
\end{split}
\end{equation}
$P^{D\bar D}$ represents the momentum of pure bound state in the rest frame, whose fourth component is different from the one of $P$. This BS wave function should satisfy the equation
\begin{equation}\label{BSE}
\chi^{0^+}_{\lambda\tau}(P^{D\bar D},p)=-\int \frac{d^4p'}{(2\pi)^4}\Delta_{F\lambda\theta}(p_1')\mathcal{V}_{\theta\theta',\kappa'\kappa}(p,p';P^{D\bar D})\chi^{0^+}_{\theta'\kappa'}(P^{D\bar D},p')\Delta_{F\kappa\tau}(p_2'),
\end{equation}
where $\mathcal{V}_{\theta\theta',\kappa'\kappa}$ is the interaction kernel, $P^{D\bar D}=(0,0,0,iM_{D\bar D})$, $p_1'=p+P^{D\bar D}/2$, $p_2'=p-P^{D\bar D}/2$, $\Delta_{F\lambda\theta}(p_1')$ and $\Delta_{F\kappa\tau}(p_2')$ are the propagators for the spin 1 fields, $\Delta_{F\lambda\theta}(p_1')=(\delta_{\lambda\theta}+\frac{p'_{1\lambda} p'_{1\theta}}{M_1^2})\frac{-i}{p_1'^2+M_1^2-i\varepsilon}$, $\Delta_{F\kappa\tau}(p_2')=(\delta_{\kappa\tau}+\frac{p'_{2\kappa} p'_{2\tau}}{M_2^2})\frac{-i}{p_2'^2+M_2^2-i\varepsilon}$, $M_1=M_{D^*_l}$ and $M_2=M_{\bar{D}_l^{*}}$. We emphasize that the kernel $\mathcal{V}$ is defined in two-body channel so $\mathcal{V}$ is not complete interaction. The kernel in homogeneous BS equation (\ref{BSE}) plays a central role for making two-body system to be a stable bound state, and the solution of homogeneous BS equation (\ref{BSE}) should only describe bound state.

\begin{figure}[!htb] \centering
\includegraphics[trim = 0mm 50mm 0mm 30mm,scale=1,width=10cm]{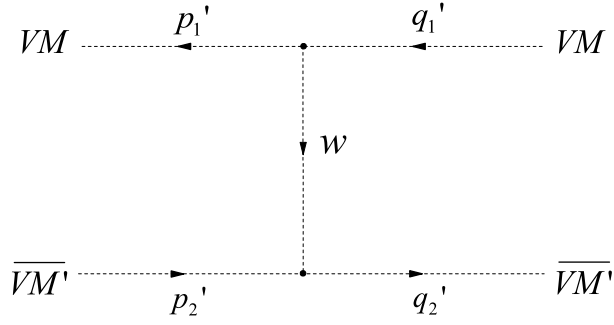}
\caption{\label{Fig1} The light meson exchange between two heavy vector mesons.}
\end{figure}

To construct the interaction kernel between $D^{*}_l$ and $\bar D^{*}_l$, we consider that the effective interaction is derived from one light meson ($\sigma$, $\rho^0$, $V_1$ and $V_8$) exchange \cite{mypaper3,mypaper4,mypaper5}, shown as Fig. \ref{Fig1}. The flavor-SU(3) singlet $V_1$ and octet $V_8$ states of vector mesons mix to form the physical $\omega$ and $\phi$ mesons as
\begin{equation}\label{mix}
\phi=-V_8\cos\theta+V_1\sin\theta,\quad \omega=V_8\sin\theta+V_1\cos\theta,
\end{equation}
where the mixing angle $\theta=38.58^\circ$ was obtained by KLOE \cite{mix}. Then the exchanged mesons should be the octet $V_8$ and singlet $V_1$ states, and the relation of the octet-quark coupling constant $g_8$ and the singlet-quark coupling constant $g_1$ has the form
\begin{equation}
g_\phi=-g_8\cos\theta+g_1\sin\theta,\quad g_\omega=g_8\sin\theta+g_1\cos\theta,
\end{equation}
where the meson-quark coupling constants $g_\omega^2=2.42/2$ and $g_\phi^2=13.0$ were determined by QCD sum rules approach \cite{cc2}.

In Fig. \ref{Fig1}, $VM$ represents $D_l^{*}$ and $\overline{VM'}$ represents $\bar D_l^{*}$. From the Lorentz-structure, the matrix elements of quark scalar density $J$ and quark current $J_\alpha$ can be expressed as
\begin{subequations}\label{vertice}
\begin{equation}
\begin{split}
\langle &VM^\varrho(p_1')|J|VM^\vartheta(q_1') \rangle=\frac{1}{2\sqrt{E_{D^*_l}(p_1')E_{D^*_l}(q_1')}}\\
&\times\bigg\{[\varepsilon^\varrho(p_1')\cdot\varepsilon^\vartheta(q_1')]h^{(\text{s})}_{1}(w^{2})-h^{(\text{s})}_{2}(w^{2})\frac{1}{M_{1}^{2}}[\varepsilon^\varrho(p_1')\cdot q_1'][\varepsilon^\vartheta(q_1')\cdot p_1']\bigg\},
\end{split}
\end{equation}
\begin{equation}
\begin{split}
\langle &\overline{VM'}^{\varrho'}(-p_2')|J|\overline{VM'}^{\vartheta'}(-q_2')\rangle=\frac{1}{2\sqrt{E_{\bar D^*_l}(-p_2')E_{\bar D^*_l}(-q_2')}}\\
&\times\bigg\{[\varepsilon^{\varrho'}(-p_2')\cdot\varepsilon^{\vartheta'}(-q_2')]\bar{h}^{(\text{s})}_{1}(w^{2})-\bar{h}^{(\text{s})}_{2}(w^{2})\frac{1}{M_{2}^{2}}[\varepsilon^{\varrho'}(-p_2')\cdot(-q_2')][\varepsilon^{\vartheta'}(-q_2')\cdot (-p_2')]\bigg\},
\end{split}
\end{equation}
\begin{equation}
\begin{split}
\langle &VM^\varrho(p_1')|J_{\alpha}|VM^\vartheta(q_1')\rangle=\frac{1}{2\sqrt{E_{D^*_l}(p_1')E_{D^*_l}(q_1')}}\\
&\times\bigg\{[\varepsilon^\varrho(p_1')\cdot\varepsilon^\vartheta(q_1')]h_{1}^{(\text{v})}(w^{2})(p_1'+q_1')_{\alpha}-h^{(\text{v})}_{2}(w^{2})\{[\varepsilon^\varrho(p_1')\cdot q_1']\varepsilon^\vartheta_{\alpha}(q_1')\\
&+[\varepsilon^\vartheta(q_1')\cdot p_1']\varepsilon^\varrho_{\alpha}(p_1')\}-h_{3}^{(\text{v})}(w^{2})\frac{1}{M_1^{2}}[\varepsilon^\varrho(p_1')\cdot q_1'][\varepsilon^\vartheta(q_1')\cdot p_1'](p_1'+q_1')_{\alpha}\bigg\},
\end{split}
\end{equation}
\begin{equation}
\begin{split}
\langle &\overline{VM'}^{\varrho'}(-p_2')|J_{\beta}|\overline{VM'}^{\vartheta'}(-q_2')\rangle=\frac{1}{2\sqrt{E_{\bar D^*_l}(-p_2')E_{\bar D^*_l}(-q_2')}}\\
&\times\bigg\{[\varepsilon^{\varrho'}(-p_2')\cdot\varepsilon^{\vartheta'}(-q_2')]\bar{h}_{1}^{(\text{v})}(w^{2})(-p_2'-q_2')_{\beta}\\
&-\bar{h}^{(\text{v})}_{2}(w^{2})\{[\varepsilon^{\varrho'}(-p_2')\cdot (-q_2')]\varepsilon^{\vartheta'}_{\beta}(-q_2')+[\varepsilon^{\vartheta'}(-q_2')\cdot (-p_2')]\varepsilon^{\varrho'}_{\beta}(-p_2')\}\\
&-\bar{h}_{3}^{(\text{v})}(w^{2})\frac{1}{M_2^{2}}[\varepsilon^{\varrho'}(-p_2')\cdot (-q_2')][\varepsilon^{\vartheta'}(-q_2')\cdot (-p_2')](-p_2'-q_2')_{\beta}\bigg\},
\end{split}
\end{equation}
\end{subequations}
where $p'_1=(\mathbf{p},ip_{10}')$, $p'_2=(\mathbf{p},ip_{20}')$, $q'_1=(\mathbf{p}',iq_{10}')$, $q'_2=(\mathbf{p}',iq_{20}')$, $w=q_1'-p_1'=q_2'-p_2'$ is the momentum of the light meson and $\mathbf{w}=\mathbf{p}'-\mathbf{p}$; $h(w^2)$ and $\bar h(w^2)$ are scalar functions, the four-vector $\varepsilon(p)$ is the polarization vector of heavy vector meson with momentum $p$ and $E_{D_l^*}(p)=\sqrt{\mathbf{p}^{2}+M_{D_l^*}^{2}}$. Taking away the external lines including normalizations and polarization vectors $\varepsilon^\varrho_\theta(p_1')$, $\varepsilon^\vartheta_{\theta'}(q_1')$, $\varepsilon^{\varrho'}_\kappa(-p_2')$, $\varepsilon^{\vartheta'}_{\kappa'}(-q_2')$, we obtain the interaction kernel from one light meson ($\sigma$, $\rho^0$, $V_1$ and $V_8$) exchange \cite{mypaper3,mypaper4}
\begin{equation}\label{kernel}
\begin{split}
\mathcal {V}&_{\theta\theta',\kappa'\kappa}(p,p';P^{D\bar D})\\
=&h_1^{(\text{s})}(w^2)\frac{-ig_\sigma^2}{w^2+M_\sigma^2}\bar{h}_1^{(\text{s})}(w^2)\delta_{\theta\theta'}\delta_{\kappa'\kappa}+\left(\frac{-ig_\rho^2}{w^2+M_\rho^2}+\frac{-ig_1^2}{w^2+M_\omega^2}+\frac{-ig_8^2}{w^2+M_\phi^2}\right)\{h_1^{(\text{v})}(w^2)\bar{h}_1^{(\text{v})}(w^2)\\
&\times(p_1'+q_1')\cdot(-p_2'-q_2')\delta_{\theta\theta'}\delta_{\kappa'\kappa}-h_1^{(\text{v})}(w^2)\bar{h}_2^{(\text{v})}(w^2)\delta_{\theta\theta'}[-(p_1'+q_1')_{\kappa'} q_{2\kappa}'-p_{2\kappa'}'(p_1'+q_1')_\kappa]\\
&-h_2^{(\text{v})}(w^2)\bar{h}_1^{(\text{v})}(w^2)[q_{1\theta}'(-p_2'-q_2')_{\theta'}+(-p_2'-q_2')_\theta p_{1\theta'}']\delta_{\kappa'\kappa}+h_2^{(\text{v})}(w^2)\bar{h}_2^{(\text{v})}(w^2)[-q_{1\theta}'\delta_{\theta'\kappa'}q_{2\kappa}'\\
&+q_{1\theta}'\delta_{\theta'\kappa}(-p_{2\kappa'}')-\delta_{\theta\kappa'}p_{1\theta'}'q_{2\kappa}'+\delta_{\theta\kappa}p_{1\theta'}'(-p_{2\kappa'}')]\},
\end{split}
\end{equation}
where $g_\sigma=\frac{B(M_\sigma)}{f_\sigma}=\frac{299}{60}$ \cite{cc1a,cc1b}, $g_\rho^2=2.42$ \cite{cc2}, and these terms containing $M_{1,2}$ are neglected because the masses of heavy mesons are large. Using the method above, we can obtain the interaction kernels from one-$\rho^{\pm}$ exchange \cite{mypaper5}.

\subsection{Instantaneous approximation}\label{sec:Instanapp}
\subsubsection{The extended Bethe-Salpeter equation}\label{sec:extenedBSE}
Substituting BS wave function given by Eq. (\ref{bswf0+}) and the kernel (\ref{kernel}) into BS equation (\ref{BSE}), we find that the integral of one term on the right-hand side of (\ref{bswf0+}) has contribution to the one of itself and the other term. Ignoring the cross terms, one can obtain two individual equations:
\begin{equation}\label{BSE1}
\mathcal{F}^1_{\lambda\tau}(P^{D\bar D}\cdot p,p^2)=-\int\frac{d^4p'}{(2\pi)^4}\Delta_{F\lambda\theta}(p_1')\mathcal{V}_{\theta\theta',\kappa'\kappa}(p,p';P^{D\bar D})\mathcal{F}^1_{\theta'\kappa'}(P^{D\bar D}\cdot p',p'^2)\Delta_{F\kappa\tau}(p_2'),
\end{equation}
\begin{equation}\label{BSE2}
\mathcal{F}^2_{\lambda\tau}(P^{D\bar D}\cdot p,p^2)=-\int\frac{d^4p'}{(2\pi)^4}\Delta_{F\lambda\theta}(p_1')\mathcal{V}_{\theta\theta',\kappa'\kappa}(p,p';P^{D\bar D})\mathcal{F}^2_{\theta'\kappa'}(P^{D\bar D}\cdot p',p'^2)\Delta_{F\kappa\tau}(p_2'),
\end{equation}
where $\mathcal{F}^1_{\lambda\tau}(P^{D\bar D}\cdot p,p^2)=\mathcal{T}^1_{\lambda\tau}\mathcal{F}_1(P^{D\bar D}\cdot p,p^2)$ and $\mathcal{F}^2_{\lambda\tau}(P^{D\bar D}\cdot p,p^2)=\mathcal{T}^2_{\lambda\tau}\mathcal{F}_2(P^{D\bar D}\cdot p,p^2)$. Comparing the tensor structures in both sides of Eqs. (\ref{BSE1}) and (\ref{BSE2}), respectively, we obtain
\begin{equation}\label{BSEF1}
\begin{split}
\mathcal{F}_1(P^{D\bar D}\cdot p,p^2)=\frac{1}{p_1'^2+M_1^2-i\varepsilon}\frac{1}{p_2'^2+M_2^2-i\varepsilon}\int\frac{d^4p'}{(2\pi)^4}V^{0^+}_{1}(p,p';P^{D\bar D})\mathcal{F}_1(P^{D\bar D}\cdot p',p'^2),
\end{split}
\end{equation}
\begin{equation}\label{BSEF2}
\begin{split}
p_2'^2\mathcal{F}_2(P^{D\bar D}\cdot p,p^2)=\frac{1}{p_1'^2+M_1^2-i\varepsilon}\frac{1}{p_2'^2+M_2^2-i\varepsilon}\int\frac{d^4p'}{(2\pi)^4}V^{0^+}_{2}(p,p';P^{D\bar D})q_2'^2\mathcal{F}_2(P^{D\bar D}\cdot p',p'^2),
\end{split}
\end{equation}
where $V^{0^+}_{1}(p,p';P^{D\bar D})$ and $V^{0^+}_{2}(p,p';P^{D\bar D})$ are derived from the interaction kernel between $D_l^{*}$ and $\bar{D}_l^{*}$. In instantaneous approximation, we set the momentum of exchanged meson as $w=(\mathbf{w},0)$. Then Eqs. (\ref{BSEF1}) and (\ref{BSEF2}) become two relativistic Schr$\ddot{o}$dinger-like equations (see details in Refs. \cite{mypaper4,mypaper5})
\begin{equation}\label{BSE3}
\left(\frac{b_1^2(M_{D\bar D})}{2\mu_R}-\frac{\mathbf{p}^2}{2\mu_R}\right)\Psi_1^{0^+}(\mathbf{p})=\int\frac{d^3w}{(2\pi)^3}V_{1}^{0^+}(\mathbf{p},\mathbf{w})\Psi_1^{0^+}(\mathbf{p},\mathbf{w}),
\end{equation}
\begin{equation}\label{BSE4}
\left(\frac{b_2^2(M_{D\bar D})}{2\mu_R}-\frac{\mathbf{p}^2}{2\mu_R}\right)\Psi_2^{0^+}(\mathbf{p})=\int\frac{d^3w}{(2\pi)^3}V_{2}^{0^+}(\mathbf{p},\mathbf{w})\Psi_2^{0^+}(\mathbf{p},\mathbf{w}),
\end{equation}
where $\Psi_1^{0^+}(\mathbf{p})=\int dp_0\mathcal{F}_1(P^{D\bar D}\cdot p,p^2)$, $\Psi_2^{0^+}(\mathbf{p})=\int dp_0p_2'^2\mathcal{F}_2(P^{D\bar D}\cdot p,p^2)$, $\mu_R=E_1E_2/(E_1+E_2)=[M_{D\bar D}^4-(M_1^2-M_2^2)^2]/(4M_{D\bar D}^3)$, $b^2(M_{D\bar D})=[M_{D\bar D}^2-(M_1+M_2)^2][M_{D\bar D}^2-(M_1-M_2)^2]/(4M_{D\bar D}^2)$, $E_1=(M_{D\bar D}^2-M_2^2+M_1^2)/(2M_{D\bar D})$ and $E_2=(M_{D\bar D}^2-M_1^2+M_2^2)/(2M_{D\bar D})$. The potentials between $D_l^{*}$ and $\bar{D}_l^{*}$ up to the second order of the $p/M_{D^*_l}$ expansion are
\begin{equation}\label{potential1}
\begin{split}
V_{1}^{0^+}(\mathbf{p},\mathbf{w})=&\frac{h_1^{(\text{s})}(w^2)}{2E_1}\frac{g_\sigma^2}{w^2+M_\sigma^2}\frac{\bar{h}_1^{(\text{s})}(w^2)}{2E_2}+h_1^{(\text{v})}(w^2)\bar{h}_1^{(\text{v})}(w^2)\\
&\times\bigg(\frac{g_\rho^2}{w^2+M_\rho^2}+\frac{g_1^2}{w^2+M_\omega^2}+\frac{g_8^2}{w^2+M_\phi^2}\bigg)\left(-1-\frac{4\mathbf{p}^2+5\mathbf{w}^2}{4E_1E_2}\right),
\end{split}
\end{equation}
\begin{equation}\label{potential2}
\begin{split}
V_{2}^{0^+}(\mathbf{p},\mathbf{w})=&\frac{h_1^{(\text{s})}(w^2)}{2E_1}\frac{g_\sigma^2}{w^2+M_\sigma^2}\frac{\bar{h}_1^{(\text{s})}(w^2)}{2E_2}\left(1-\frac{\mathbf{w}^2}{M_1^2}\right)+h_1^{(\text{v})}(w^2)\bar{h}_1^{(\text{v})}(w^2)\bigg(\frac{g_\rho^2}{w^2+M_\rho^2}\\
&+\frac{g_1^2}{w^2+M_\omega^2}+\frac{g_8^2}{w^2+M_\phi^2}\bigg)\left(-1-\frac{2\mathbf{p}^2+2\mathbf{w}^2}{4M_1^2}-\frac{2\mathbf{p}^2+2\mathbf{w}^2}{4E_1E_2}\right).
\end{split}
\end{equation}
Solving these two equations (\ref{BSE3}) and (\ref{BSE4}), respectively, one can obtain the eigenvalues $b_1^2(M_{D\bar D})$ and $b_2^2(M_{D\bar D})$ and the corresponding eigenfunctions $\Psi_1^{0^+}(\mathbf{p})$ and $\Psi_2^{0^+}(\mathbf{p})$. From $\Psi_1^{0^+}$ and $\Psi_2^{0^+}$, it is easy to obtain $\mathcal{F}_1$ and $\mathcal{F}_2$, respectively.

Because the cross terms are small, we can take the ground state BS wave function to be a linear combination of two eigenstates $\mathcal{F}^{10}_{\lambda\tau}(P^{D\bar D}\cdot p,p^2)$ and $\mathcal{F}^{20}_{\lambda\tau}(P^{D\bar D}\cdot p,p^2)$ corresponding to lowest energy in Eqs. (\ref{BSE1}) and (\ref{BSE2}). Then in the basis provided by $\mathcal{F}^{10}_{\lambda\tau}(P^{D\bar D}\cdot p,p^2)=\mathcal{T}^1_{\lambda\tau}\mathcal{F}_{10}(P^{D\bar D}\cdot p,p^2)$ and $\mathcal{F}^{20}_{\lambda\tau}(P^{D\bar D}\cdot p,p^2)=\mathcal{T}^2_{\lambda\tau}\mathcal{F}_{20}(P^{D\bar D}\cdot p,p^2)$, BS wave function $\chi^{0^+}_{\lambda\tau}$ is considered as
\begin{equation}\label{BSwfapprox}
\chi^{0^+}_{\lambda\tau}(P^{D\bar D},p)=\frac{1}{\mathcal{N}_{D\bar D}^{0^+}}[\mathcal{C}_1\mathcal{F}^{10}_{\lambda\tau}(P^{D\bar D}\cdot p,p^2)+\mathcal{C}_2\mathcal{F}^{20}_{\lambda\tau}(P^{D\bar D}\cdot p,p^2)].
\end{equation}
Substituting (\ref{BSwfapprox}) into BS equation (\ref{BSE}) and comparing the tensor structures in both sides, we obtain an eigenvalue equation in instantaneous approximation \cite{mypaper4}
\begin{equation}\label{eigeneq}
\left(\begin{array}{cc}\frac{b_{10}^2(M_{D\bar D})}{2\mu_R}-\lambda&H_{12}\\H_{21}&\frac{b_{20}^2(M_{D\bar D})}{2\mu_R}-\lambda\end{array}\right)\left(\begin{array}{c}\mathcal{C}'_1\\\mathcal{C}'_2\end{array}\right)=0,
\end{equation}
where we have the matrix elements
\begin{equation}
\begin{split}
H_{12}=H_{21}=&\int \frac{d^3p}{(2\pi)^3}\Psi_{10}^{0^+}(\mathbf{p})^*\int\frac{d^3w}{(2\pi)^3}h_1^{(\text{v})}(w^2)\\
&\times\left(\frac{g_\rho^2}{w^2+M_\rho^2}+\frac{g_1^2}{w^2+M_\omega^2}+\frac{g_8^2}{w^2+M_\phi^2}\right)\bar{h}_1^{(\text{v})}(w^2)\frac{\mathbf{w}^2}{E_1E_2}\Psi_{20}^{0^+}(\mathbf{p},\mathbf{w}),
\end{split}
\end{equation}
and $b_{10}^2(M_{D\bar D})/(2\mu_R)$ and $b_{20}^2(M_{D\bar D})/(2\mu_R)$ are the eigenvalues corresponding to lowest energy in Eqs. (\ref{BSE3}) and (\ref{BSE4}), respectively; $\Psi_{10}^{0^+}$ and $\Psi_{20}^{0^+}$ are the corresponding eigenfunctions. From this equation, we can obtain the eigenvalues and eigenfunctions which contain the contribution from the cross terms. Some errors in our previous works have been revised.

\subsubsection{Form factors of heavy meson}\label{sec:Formfac}
To calculate these heavy vector meson form factors $h(w^2)$ describing the heavy meson structure, we have to know the wave function of heavy vector meson $D_l^*$ in instantaneous approximation. For heavy vector mesons, the authors of Refs. \cite{BSE:Roberts1,BSE:Roberts3,BSE:Roberts4,BSE:Roberts5} have obtained their BS amplitudes in Euclidean space:
\begin{equation}
\begin{split}
\Gamma_{\lambda}^{V}(K,k)=\frac{1}{\mathcal{N}^{V}}\bigg(\gamma_\lambda+K_{\lambda}\frac{\gamma\cdot K}{M_{V}^2}\bigg)\varphi_{V}(k^2),
\end{split}
\end{equation}
where $K$ is the momentum of heavy meson, $k$ denotes the relative momentum between quark and antiquark in heavy meson, $M_V$ is heavy vector meson mass, $\Gamma_{\lambda}^{V}(K,k)$ is transverse ($K_{\lambda}\Gamma_{\lambda}^{V}(K,k)=0$), $\mathcal{N}^V$ is normalization, and $\varphi_{V}(k^{2})$ is scalar function fixed by providing fits to observables. The charmed meson $D_l^{*}$ is composed of $c$-quark and $l$-antiquark. As in heavy-quark effective theory (HQET) \cite{hqet}, we consider that the heaviest quark carries all the heavy-meson momentum and obtain BS wave function of $D_l^*$ meson
\begin{equation}\label{D0BSwf}
\begin{split}
\chi_\lambda(K,k)=\frac{-1}{\gamma\cdot(k+K)-im_c}\frac{1}{\mathcal{N}^{D_l^{*}}}\bigg(\gamma_\lambda+K_{\lambda}\frac{\gamma\cdot K}{M_{D_l^{*}}^2}\bigg)\varphi_{D_l^{*}}(k^2)\frac{-1}{\gamma\cdot k-im_l},
\end{split}
\end{equation}
where $K$ is set as the momentum of heavy meson in the rest frame, $k$ becomes the relative momentum between $c$-quark and $l$-antiquark, $m_{c,l}$ are the constituent quark masses, $\varphi_{D_l^{*}}(k^{2})=\varphi_{\bar D_l^{*}}(k^{2})=\exp(-k^{2}/\omega_{D^{*}}^{2})$ and $\omega_{D^{*}}$=1.50 GeV \cite{BSE:Roberts5}. The components of this BS wave function are $4\times4$ matrices, which can be written as \cite{mypaper}
\begin{equation}
\begin{split}
\chi_{\lambda}(K,k)=\Psi_\lambda^{S}+\Psi^{V}_{\lambda,\mu}\gamma_{\mu}+\Psi^{T}_{\lambda,\mu\nu}\sigma_{\mu\nu}+\Psi^{AV}_{\lambda,\mu}\gamma_{\mu}\gamma_{5}+\Psi^{Pse}_\lambda\gamma_{5},
\end{split}
\end{equation}
and the coefficient corresponding to $\gamma_\mu$ is
\begin{equation}\label{BSwfTr}
\begin{split}
\Psi_{\lambda,\mu}^{V}=\frac{1}{4}\text{Tr}[\gamma_\mu\chi_\lambda(K,k)].
\end{split}
\end{equation}
Substituting Eq. (\ref{D0BSwf}) into (\ref{BSwfTr}), we can obtain the heavy vector meson wave function in instantaneous approximation
\begin{equation}\label{wfV}
\begin{split}
\Psi_{ij}^{D_l^*}(\mathbf{k})&=\int dk_4\frac{1}{N^{D_l^*}}\exp\bigg(\frac{-\mathbf{k}^2-k_4^2}{\omega_{D_l^*}^2}\bigg)\frac{\mathbf{k}^2/3+k_4^2+m_cm_l}
{(\mathbf{k}^2+k_4^2+m_c^2)(\mathbf{k}^2+k_4^2+m_l^2)}\delta_{ij}~~i,j=1,2,3.
\end{split}
\end{equation}

In the previous works \cite{mypaper3,mypaper4,mypaper5}, we have obtained the form factors for the vertices of heavy vector meson $D_l^*$ coupling to scalar meson ($\sigma$)
\begin{equation}\label{ffp}
\begin{split}
-\frac{h_{1}^{(\text{s})}(w^{2})}{2E_1}=&\frac{\bar{h}_{1}^{(\text{s})}(w^{2})}{2E_2}=F_1({\mathbf{w}^{2}}),~~~h_{2}^{(\text{s})}(w^{2})=\bar{h}_{2}^{(\text{s})}(w^{2})=0,\\
F_{1}(\mathbf{w}^{2})=&\int \frac{d^{3}k}{(2\pi)^{3}}\bar{\Psi}^{D_l^*}\bigg(\mathbf{k}+\frac{2E_c(k)}{E_{D_l^*}+M_{D_l^*}}\mathbf{w}\bigg)\sqrt{\frac{E_l(k)+m_l}{E_l(k+w)+m_l}}\\
&\times\left\{\frac{E_l(k+w)-E_l(k)+2m_l}{2\sqrt{E_l(k+w)E_l(k)}}-\frac{\mathbf{k\cdot w}}{2\sqrt{E_l(k+w)E_l(k)}[E_l(k)+m_l]}\right\}\Psi^{D_l^*}(\mathbf{k}),
\end{split}
\end{equation}
and to vector meson ($\rho$, $V_1$ and $V_8$)
\begin{equation}\label{ffv}
\begin{split}
h_{1}^{(\text{v})}(w^{2})=&h_{2}^{(\text{v})}(w^{2})=\bar{h}_{1}^{(\text{v})}(w^{2})=\bar{h}_{2}^{(\text{v})}(w^{2})=F_2({\mathbf{w}^{2}}),~~~h_{3}^{(\text{v})}(w^{2})=\bar{h}_{3}^{(\text{v})}(w^{2})=0,\\
F_2(\mathbf{w}^{2})=&\frac{2\sqrt{E_{D_l^*}M_{D_l^*}}}{E_{D_l^*}+M_{D_l^*}}\int\frac{d^{3}k}{(2\pi)^{3}}\bar{\Psi}^{D_l^*}\bigg(\mathbf{k}+\frac{2E_c(k)}{E_{D_l^*}+M_{D_l^*}}\mathbf{w}\bigg)\sqrt{\frac{E_l(k)+m_l}{E_l(k+w)+m_l}}\\
&\times\left\{\frac{E_l(k+w)+E_l(k)}{2\sqrt{E_l(k+w)E_l(k)}}+\frac{\mathbf{k\cdot w}}{2\sqrt{E_l(k+w)E_l(k)}[E_l(k)+m_l]}\right\}\Psi^{D_l^*}(\mathbf{k}),
\end{split}
\end{equation}
where $E_{c,l}(p)=\sqrt{\mathbf{p}^{2}+m_{c,l}^{2}}$ and $\Psi^{D_l^*}$ is the wave function of heavy vector meson expressed as Eq. (\ref{wfV}). In this paper, some errors in our previous works have been revised. Equations (\ref{BSE3}) and (\ref{BSE4}) can be solved numerically with these form factors, and then the eigenvalue equation (\ref{eigeneq}) can be solved. The masses $M_{D\bar D}$ and wave functions of pure bound states $D^{*0}\bar{D}^{*0}$ and $D^{*+}D^{*-}$ with spin-parity quantum numbers $0^+$ can be obtained.

Considering the interaction kernels from one-$\rho^{\pm}$ exchange and using the coupled-channel approach (see details in Ref. \cite{mypaper5}), we can calculate the mass $M_0$ of the mixed state of two pure bound states $D^{*0}\bar{D}^{*0}$ and $D^{*+}D^{*-}$ with $0^+$. Since the mixing of component wave functions causes the change of energy, the fourth component of $P^{D\bar D}$ in the original BS wave function becomes the total energy of mixed state, and $\chi_{\lambda\tau}^{0^+}(P^{D\bar D},p)$ in Eq. (\ref{BSwfapprox}) becomes
\begin{equation}\label{bswf0+1}
\begin{split}
\chi_{\lambda\tau}^{0^+}(P,p)=&\frac{1}{\mathcal{N}^{0^+}}\{[(p'_1\cdot p'_2)g_{\lambda\tau}-p'_{2\lambda}p'_{1\tau}]\mathcal{C}_1\mathcal{F}_{10}(P\cdot p,p^2)\\
&+[p_1'^2p_2'^2g_{\lambda\tau}+(p_1'\cdot p_2')p'_{1\lambda}p'_{2\tau}-p_2'^2p'_{1\lambda}p'_{1\tau}-p_1'^2p'_{2\lambda}p'_{2\tau}]\mathcal{C}_2\mathcal{F}_{20}(P\cdot p,p^2)\}.
\end{split}
\end{equation}
We emphasize that the mass $M_0$ of meson-meson bound state should not be the mass of physical resonance. Substituting Eq. (\ref{bswf0+1}) into (\ref{mmbounstate}), we obtain BS wave function $\chi^{D^{*}\bar D^{*},0^+}_{\lambda\tau}(P,p)$ for the mixed state of two bound states $D^{*0}\bar{D}^{*0}$ and $D^{*+}D^{*-}$ with $0^+$.

\subsection{GBS wave function for four-quark state}\label{sec:GBS}
\begin{figure}[!htb] \centering
\includegraphics[trim = 0mm 20mm 0mm 20mm,scale=1,width=10cm]{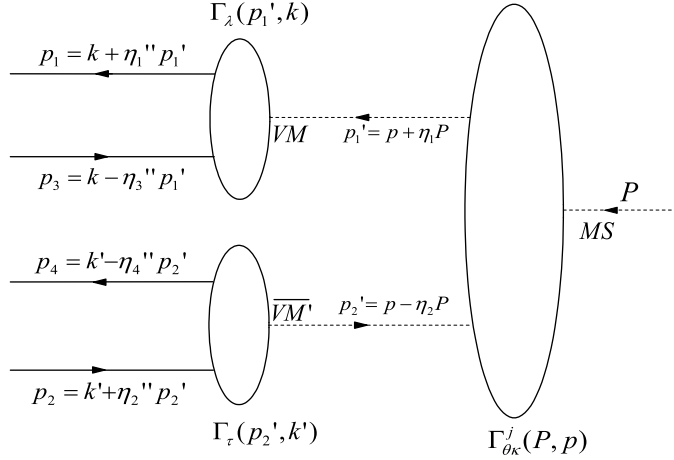}
\caption{\label{Fig2}  Generalized Bethe-Salpeter wave function for four-quark state in the momentum representation. The solid lines denote quark propagators, and the unfilled ellipses represent Bethe-Salpeter amplitudes.}
\end{figure}

The heavy meson is a bound state consisting of a quark and an antiquark and the meson-meson bound state is actually composed of four quarks. We have to give GBS wave function of meson-meson bound state as a four-quark state. If a bound state with spin $j$ and parity $\eta_{P}$ is composed of four quarks, its GBS wave function can be defined as \cite{mypaper6}
\begin{equation}\label{GBSWF}
\chi^j_P(x_1,x_3,x_4,x_2)=\langle 0|T\mathcal{Q}^\mathcal{C}(x_1)\bar{\mathcal{Q}}^\mathcal{A}(x_3)\mathcal{Q}^\mathcal{B}(x_4)\bar{\mathcal{Q}}^\mathcal{D}(x_2)|P,j\rangle,
\end{equation}
where $P$ is the momentum of the four-quark bound state, $\mathcal{Q}$ is the quark operator and its superscript is a flavor label. From translational invariance, this GBS wave function can be written as
\begin{equation}
\chi^j_{P}(x_1,x_3,x_4,x_2)=\frac{1}{(2\pi)^{3/2}}\frac{1}{\sqrt{2E(P)}}e^{iP\cdot X}\chi^j_{P}(X',x,x'),
\end{equation}
where $X=\eta_1(\eta_1''x_1+\eta_3''x_3)+\eta_2(\eta_4''x_4+\eta_2''x_2)$, $X'=(\eta_1''x_1+\eta_3''x_3)-(\eta_4''x_4+\eta_2''x_2)$, $x=x_1-x_3$, $x'=x_2-x_4$, $\eta_1+\eta_2=1$, $\eta''_{1,3}=m_{\mathcal{C},\mathcal{A}}/(m_\mathcal{C}+m_\mathcal{A})$, $\eta''_{2,4}=m_{\mathcal{D},\mathcal{B}}/(m_\mathcal{D}+m_\mathcal{B})$ and $m_{\mathcal{A},\mathcal{B},\mathcal{C},\mathcal{D}}$ are the quark masses. In the momentum representation, GBS wave function of four-quark bound state becomes
\begin{equation}\label{fourquarkBSWF}
\begin{split}
\chi^j_{P}(p_1,p_3,p_4,p_2)=\frac{1}{(2\pi)^{3/2}}\frac{1}{\sqrt{2E(P)}}(2\pi)^4\delta^{(4)}(P-p_1+p_3-p_4+p_2)\chi^j(P,p,k,k'),
\end{split}
\end{equation}
where $p_1, p_3, p_4, p_2$ are the momenta carried by the fields $\mathcal{Q}^\mathcal{C}$, $\bar{\mathcal{Q}}^\mathcal{A}$, $\mathcal{Q}^\mathcal{B}$, $\bar{\mathcal{Q}}^\mathcal{D}$; $p$, $k$, $k'$ are the conjugate variables to $X'$, $x$, $x'$, respectively; and $p=\eta_2(p_1-p_3)+\eta_1(p_2-p_4)$, $k=\eta_3''p_1+\eta_1''p_3$, $k'=\eta_4''p_2+\eta_2''p_4$. In the hadronic molecule structure, $p$ is the relative momentum between two mesons in molecular state, $k$ and $k'$ are the relative momenta between quark and antiquark in these two mesons, respectively, shown as Fig. \ref{Fig2}. This work is aimed to investigate the bound state composed of two vector mesons. In Fig. \ref{Fig2}, $VM$ represents the vector meson with mass $M_1$, $\overline{VM'}$ represents the anti-particle of vector meson $VM'$ with mass $M_2$, and $MS$ represents the meson-meson bound state.

In Fig. \ref{Fig2}, there are three two-body systems: a meson-meson bound state and two quark-antiquark bound states. We define BS wave functions of these two-body systems as $\chi^j_P(p_1',p_2')$, $\chi_{p_1'}(p_1,p_3)$, $\chi_{p_2'}(p_4,p_2)$, respectively. BS wave function for the bound state of two vector mesons has been given by Eq. (\ref{BSwfdp}) and BS wave functions of two vector mesons are
\begin{equation}
\begin{split}
\chi_{p_1'}(p_1,p_3)_\lambda=\frac{1}{(2\pi)^{3/2}}\frac{1}{\sqrt{2E(p'_1)}}(2\pi)^4\delta^{(4)}(p_1'-p_1+p_3)\chi_\lambda(p_1',k),
\end{split}
\end{equation}
\begin{equation}
\begin{split}
\chi_{p_2'}(p_4,p_2)_\tau=\frac{1}{(2\pi)^{3/2}}\frac{1}{\sqrt{2E(p'_2)}}(2\pi)^4\delta^{(4)}(p_2'+p_4-p_2)\chi_\tau(p_2',k'),
\end{split}
\end{equation}
where $p_1'$ and $p_2'$ are the momenta of two vector mesons, respectively, $p_1'=p+\eta_1P$, $p_2'=p-\eta_2P$ and $\eta_{1,2}=M_{1,2}/(M_1+M_2)$. Applying the Feynman rules and comparing with Eq. (\ref{fourquarkBSWF}), we obtain the revised GBS wave function for four-quark state describing the bound state composed of two vector mesons with arbitrary spin and definite parity \cite{mypaper6,mypaper7}
\begin{equation}\label{fourquarkBSWF1}
\begin{split}
\chi^j(P,p,k,k')=\chi_{\lambda}(p_1',k)\chi^j_{\lambda\tau}(P,p)\chi_\tau(p_2',k').
\end{split}
\end{equation}
From Eq. (\ref{D0BSwf}), we obtain BS wave functions of vector mesons
\begin{equation}\label{BSwfvm}
\begin{split}
&\chi_\lambda(p_1',k)=\frac{-1}{\gamma^\mathcal{C}\cdot p_1-im_\mathcal{C}}\frac{1}{\mathcal{N}^{V}}\bigg(\gamma_\lambda+p_{1\lambda}'\frac{\gamma\cdot p_1'}{M_{V}^2}\bigg)\varphi_{V}(k^2)\frac{-1}{\gamma^\mathcal{A}\cdot p_3-im_\mathcal{A}},\\
&\chi_\tau(p_2',k')=\frac{-1}{\gamma^\mathcal{B}\cdot p_4-im_\mathcal{B}}\frac{1}{\mathcal{N}^{\bar V'}}\bigg(\gamma_\tau+p_{2\tau}'\frac{\gamma\cdot p_2'}{M_{\bar V'}^2}\bigg)\varphi_{\bar V'}(k'^2)\frac{-1}{\gamma^\mathcal{D}\cdot p_2-im_\mathcal{D}}.
\end{split}
\end{equation}

In this section, we consider a mixed state of two bound states $D^{*0}\bar{D}^{*0}$ and $D^{*+}D^{*-}$ with spin-parity quantum numbers $0^+$. In Fig. \ref{Fig2}, $VM$ and $\overline{VM'}$ become $D_l^{*}$ and $\bar{D}_l^{*}$, respectively, and in Eq. (\ref{GBSWF}) the flavor labels $\mathcal{C}=\mathcal{D}$ and $\mathcal{A}=\mathcal{B}$ represent $c$-quark and $l$-quark, respectively. From Eqs. (\ref{mmbounstate}), (\ref{fourquarkBSWF1}) and (\ref{BSwfvm}), we obtain the GBS wave function for meson-meson bound state as a four-quark state
\begin{equation}\label{mmbounstate1}
\begin{split}
\chi^{D^{*}\bar D^{*},0^+}(P,p,k,k')=&\frac{1}{\sqrt{2}}\chi^{D^{*0}}_\lambda(p_1',k)\chi^{D^{*0}\bar D^{*0},0^+}_{\lambda\tau}(P,p)\chi^{\bar D^{*0}}_\tau(p_2',k')\\
&+\frac{1}{\sqrt{2}}\chi^{D^{*+}}_\lambda(p_1',k)\chi^{D^{*+}D^{*-},0^+}_{\lambda\tau}(P,p)\chi^{D^{*-}}_\tau(p_2',k'),
\end{split}
\end{equation}
where
\begin{equation}\label{BSwfvmD}
\begin{split}
&\chi^{D_l^{*}}_\lambda(p_1',k)=\frac{-1}{\gamma\cdot p_1-im_c}\frac{1}{\mathcal{N}^{D_l^{*}}}\bigg(\gamma_\lambda+p_{1\lambda}'\frac{\gamma\cdot p_1'}{M_{D_l^{*}}^2}\bigg)\varphi_{D_l^{*}}(k^2)\frac{-1}{\gamma\cdot p_3-im_l},\\
&\chi^{\bar{D}_l^{*}}_\tau(p_2',k')=\frac{-1}{\gamma\cdot p_4-im_l}\frac{1}{\mathcal{N}^{\bar{D}_l^{*}}}\bigg(\gamma_\tau+p_{2\tau}'\frac{\gamma\cdot p_2'}{M_{\bar{D}_l^{*}}^2}\bigg)\varphi_{\bar{D}_l^{*}}(k'^2)\frac{-1}{\gamma\cdot p_2-im_c}.
\end{split}
\end{equation}

\subsection{Normalization of BS wave function}\label{sec:Normal}
\subsubsection{Heavy vector meson}\label{sec:heavyvm}
Here, we determine normalizations $\mathcal{N}^{D^{*0}}$ and $\mathcal{N}^{D^{*+}}$. The authors of Refs. \cite{BSE:Roberts4,BSE:Roberts5} employed the ladder approximation to solve the BS equation for quark-antiquark state, and the reduced normalization condition for the BS wave function of $D_l^{*}$ meson given by Eq. (\ref{D0BSwf}) is
\begin{equation}
\begin{split}
\frac{-i}{(2\pi)^4}\frac{1}{3}\int d^4k\bar\chi_{\lambda}(K,k)\frac{\partial}{\partial K_0}[S_F(k+K)^{-1}]S_F(k)^{-1}\chi_\lambda(K,k)=(2K_0)^2,
\end{split}
\end{equation}
where $S_F(p)^{-1}$ is the inverse propagator for quark field and the factor $1/3$ appears because of the sum of three transverse directions.

\subsubsection{Molecular state}\label{sec:moles}
The reduced normalization condition for $\chi_{\lambda\tau}^{0^+}(P,p)$ expressed as Eq. (\ref{bswf0+1}) is
\begin{equation}
\begin{split}
&\frac{-i}{(2\pi)^4}\int d^4p\bar\chi_{\tau'\lambda'}(P,p)\frac{\partial}{\partial P_0}[\Delta_{F\lambda'\lambda}(p+P/2)^{-1}\Delta_{F\tau\tau'}(p-P/2)^{-1}]\chi_{\lambda\tau}(P,p)=(2P_0)^2,
\end{split}
\end{equation}
where $\Delta_{F\beta\alpha'}(p)^{-1}$ is the inverse propagator for the vector field with mass $m$, $\Delta_{F\beta\alpha'}(p)^{-1}=i(\delta_{\beta\alpha'}-\frac{p_{\beta}p_{\alpha'}}{p^2+m^2})(p^2+m^2)$ \cite{mypaper6}. After determining normalization $\mathcal{N}^{0^+}$, we automatically obtain the normalized BS wave function for the mixed state of two components $D^{*0}\bar{D}^{*0}$ and $D^{*+}D^{*-}$ given by Eq. (\ref{mmbounstate}). Immediately, the normalized GBS wave function for meson-meson bound state as a four-quark state expressed as Eq. (\ref{mmbounstate1}) is obtained.

\section{scattering matrix element from four-quark state to final state}\label{sec:Sme}
In experiments two strong decay modes of $\chi_{c0}(3915)$ have been observed: $J/\psi\omega$ and $D^+D^-$. The narrow state $\chi_{c0}(3915)$ was discovered in 2005 \cite{Y39401} by the Belle collaboration and for a long time a series of experiments \cite{Y39402,X39151,Y39404,X39153} only observed one strong decay mode of $\chi_{c0}(3915)$: $J/\psi\omega$ denoted as $c'_1$. In 2020 the LHCb Collaboration observed another decay channel $D^+D^-$ \cite{X39152} denoted as $c'_2$. Though the neutral channel $D^0\bar D^0$ still has not been observed, this neutral channel should exist for the isospin conservation, which is denoted as $c'_3$. In this section, we present the traditional technique to calculate decay width for these processes and revise some errors in previous works \cite{mypaper6,mypaper7}.

\subsection{Decay channel $J/\psi\omega$ with respect to mass of bound state}\label{sec:Jpsiobm}
\begin{figure}[!htb] \centering
\includegraphics[trim = 0mm 30mm 0mm 20mm,scale=1,width=10cm]{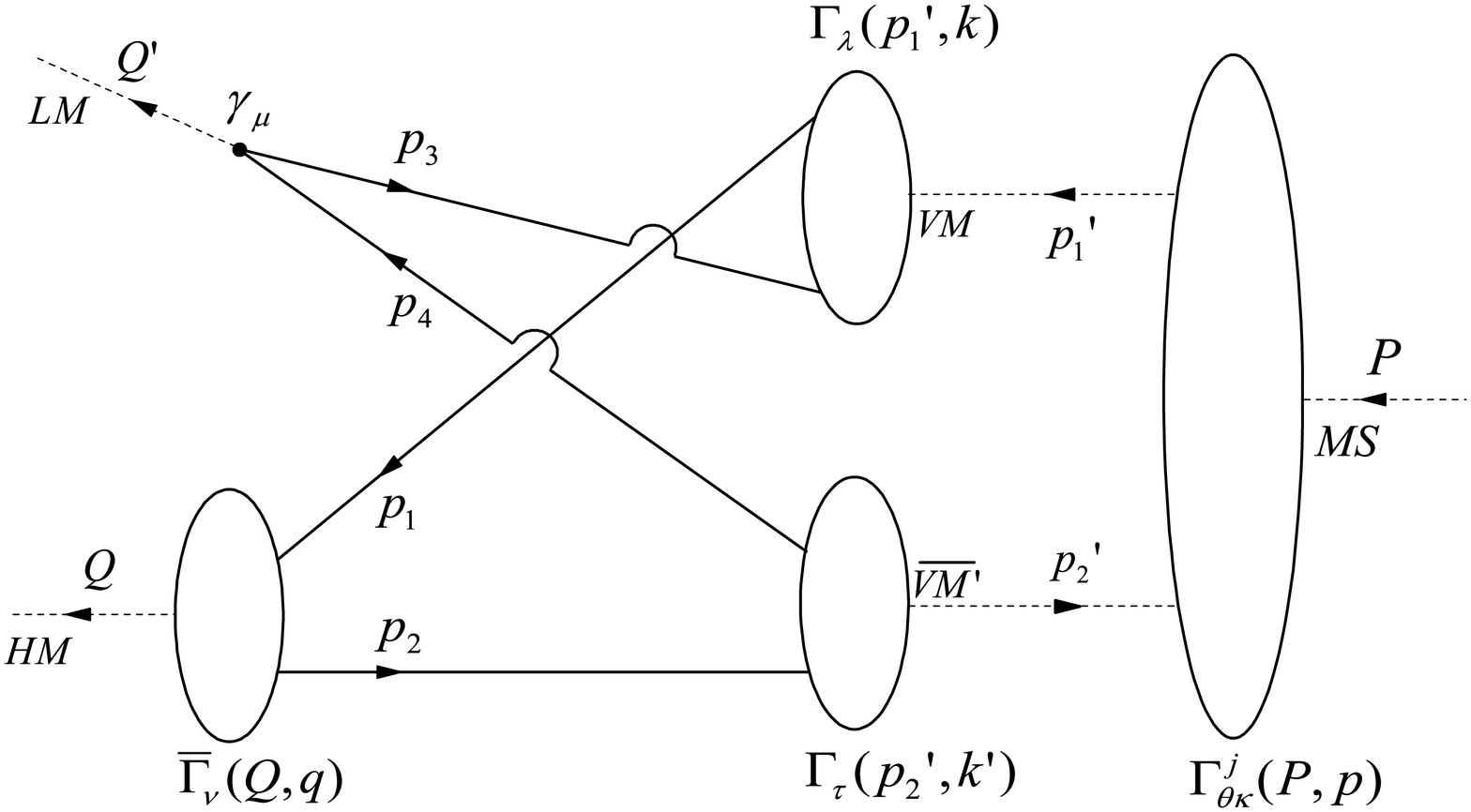}
\caption{\label{Fig3} The lowest order matrix element between bound states in the momentum representation.}
\end{figure}

Mandelstam's approach is a technique based on BS wave function for evaluating the general matrix element between bound states \cite{ParticlesFields}. Applying Mandelstam's approach, we have obtained the scattering matrix element from a four-quark state to a heavy meson plus a light meson \cite{mypaper6} in the momentum representation, as shown in Fig. \ref{Fig3}. In this work, we retain only the lowest order term of the two-particle irreducible Green's function. In Fig. \ref{Fig3}, $VM$ and $\overline{VM'}$ still represent $D_l^{*}$ and $\bar{D}_l^{*}$, respectively; $HM$ represents $J/\psi$ with momentum $Q$ ($Q^2=-M_{J/\psi}^2$) and $LM$ represents $\omega$ with momentum $Q'$ ($Q'^2=-M_{\omega}^2$). The momentum of the initial state is set as $P=(0,0,0,iM_0)$ in the rest frame, and $M_0$ is the mass of the mixed state of two pure bound states $D^{*0}\bar{D}^{*0}$ and $D^{*+}D^{*-}$, which should not be the physical mass of resonance. It is necessary to emphasize that the momenta in the final state satisfy $Q+Q'=P$ in this section. Here, we consider that in the final state the light vector meson $\omega$ is an elementary particle and the heavy vector meson $J/\psi$ is a bound state of $c\bar c$. From Eq. (\ref{BSwfvm}), we obtain the BS wave function of heavy vector meson $J/\psi$
\begin{equation}\label{jpsiBSwf}
\begin{split}
\chi_\nu(Q,q)=\frac{-1}{\gamma\cdot (q+Q/2)-im_c}\frac{1}{\mathcal{N}^{J/\psi}}\bigg(\gamma_\nu+Q_\nu\frac{\gamma\cdot Q}{M_{J/\psi}^2}\bigg)\varphi_{J/\psi}(q^2)\frac{-1}{\gamma\cdot(q-Q/2)-im_c},
\end{split}
\end{equation}
where $\varphi_{J/\psi}(q^{2})=\exp(-q^{2}/\omega_{J/\psi}^{2})$ and $\omega_{J/\psi}$=0.826 GeV was obtained from lattice QCD (see details in Ref. \cite{mypaper6}). The reduced normalization condition for BS wave function of $J/\psi$ meson expressed as Eq. (\ref{jpsiBSwf}) is
\begin{equation}
\begin{split}
\frac{-i}{(2\pi)^4}\frac{1}{3}\int d^4q\bar\chi_{\nu}(Q,q)\frac{\partial}{\partial Q_0}[S_F(q+Q/2)^{-1}S_F(q-Q/2)^{-1}]\chi_\nu(Q,q)=(2Q_0)^2,
\end{split}
\end{equation}
where the factor $1/3$ appears for the three transverse directions are summed. Normalization $\mathcal{N}^{J/\psi}$ can be determined. These momenta in Fig. \ref{Fig3} become
\begin{equation}\label{momenta}
\begin{split}
&p_1=(Q+Q')/2+p+k,~~p_2=(Q+Q')/2-Q+p+k,~~p_3=k,~~p_4=Q'+k,\\
&q=Q'/2+p+k,~~k'=Q'+k,~~p_1'=p+P/2,~~p_2'=p-P/2,~~Q+Q'=P.
\end{split}
\end{equation}

Using the Heisenberg picture, we obtain the total matrix element from the initial state $|P~\text{in}\rangle$ to a final state $\langle Q,Q'~\text{out}|$
\begin{equation}\label{Smatrele}
\begin{split}
-iR_{(c'_1;b)a}(M_0)=\langle Q,Q'~\text{out}|P~\text{in}\rangle=-i(2\pi)^4\delta^{(4)}(Q+Q'-P)T_{(c'_1;b)a}(M_0),
\end{split}
\end{equation}
where $T_{(c'_1;b)a}(M_0)$ is the $T$-matrix element with mass $M_0$ for channel $c'_1$. According to Mandelstam's approach, we obtain
\begin{equation}\label{Tmatrele}
\begin{split}
T_{(c'_1;b)a}(M_0)=\frac{ig_\omega\varepsilon_\mu^{\varrho'}(Q')\varepsilon_\nu^{\varrho}(Q)}{(2\pi)^{9/2}\sqrt{2E_{J/\psi}(Q)}\sqrt{2E_\omega(Q')}\sqrt{2E(P)}}\bigg(\frac{1}{\sqrt{2}}\mathcal{M}^{c'_1,D^{*0}\bar D^{*0}}_{\nu\mu}+\frac{1}{\sqrt{2}}\mathcal{M}^{c'_1,D^{*+}D^{*-}}_{\nu\mu}\bigg),
\end{split}
\end{equation}
where $\varepsilon_\nu^{\varrho=1,2,3}(Q)$ and $\varepsilon_\mu^{\varrho'=1,2,3}(Q')$ are the polarization vectors of $J/\psi$ and $\omega$, respectively, $\varepsilon^\varrho(Q)\cdot Q=\varepsilon^{\varrho'}(Q')\cdot Q'=0$, and
\begin{equation}\label{MEcurr}
\begin{split}
\mathcal{M}^{c'_1,D^{*}_l\bar D^{*}_l}_{\nu\mu}=&\int \frac{d^4kd^4p}{(2\pi)^8}\frac{1}{N^{J/\psi}}\frac{\varphi_{J/\psi}(q^2)}{p_2^2+m^2_c}\frac{1}{p_1^2+m^2_c}\frac{1}{N^{D_l^*}}\frac{\varphi_{D_l^*}(k^2)}{p_3^2+m^2_l}\frac{1}{N^{\bar D_l^*}}\frac{\varphi_{\bar D_l^*}(k'^2)}{p_4^2+m^2_l}\\
&\times \text{Tr}[(\gamma\cdot p_2+im_c)\gamma_\nu(\gamma\cdot p_1+im_c)\gamma_\lambda(\gamma\cdot p_3+im_l)\gamma_\mu(\gamma\cdot p_4+im_l)\gamma_\tau\chi_{\lambda\tau}^{0^+}(P,p)].
\end{split}
\end{equation}
Here $\chi_{\lambda\tau}^{0^+}(P,p)$ is expressed as Eq. (\ref{bswf0+1}). In Eq. (\ref{MEcurr}) the trace of the product of 8 $\gamma$-matrices contains 105 terms and the resulting expression has been given in Appendix B of Ref. \cite{mypaper6}. In our approach, the $p$ integral is computed in instantaneous approximation. To calculate this tensor $\mathcal{M}^{c'_1,D_l^{*}\bar{D}_l^{*}}_{\nu\mu}$, we have given a simple method in Ref. \cite{mypaper6}. It is obvious that the tensor $\mathcal{M}^{c'_1,D_l^{*}\bar{D}_l^{*}}_{\nu\mu}$ only depends on $Q$ and $Q'$, so in Minkowski space $\mathcal{M}^{c'_1,D_l^{*}\bar{D}_l^{*}}_{\nu\mu}$ can be expressed as
\begin{equation}\label{MEcurrLT}
\begin{split}
&\mathcal{M}^{c'_1,D_l^{*}\bar{D}_l^{*}}_{\nu\mu}\\
&=g_{\nu\mu}U_1(Q',Q)+Q'_\nu Q_\mu U_2(Q',Q)+Q'_\nu Q'_\mu U_3(Q',Q)+Q_\nu Q'_\mu U_4(Q',Q)+Q_\nu Q_\mu U_5(Q',Q),
\end{split}
\end{equation}
where $U_i(Q',Q)(i=1,\cdots,5)$ are scalar functions. The above expression is multiplied by these tensor structures $g_{\nu\mu}$, $Q'_\nu Q_\mu$, $Q'_\nu Q'_\mu$, $Q_\nu Q'_\mu$, $Q_\nu Q_\mu$, respectively; and a set of equations is obtained
\begin{equation}
\begin{split}
g_{\nu\mu}\mathcal{M}^{c'_1,D_l^{*}\bar{D}_l^{*}}_{\nu\mu}&=U_1'=4U_1+(Q'\cdot Q)U_2+Q'^2U_3+(Q'\cdot Q)U_4+Q^2U_5,\\
Q'_\nu Q_\mu\mathcal{M}^{c'_1,D_l^{*}\bar{D}_l^{*}}_{\nu\mu}&=U_2'=(Q'\cdot Q)U_1+Q'^2Q^2U_2+Q'^2(Q'\cdot Q)U_3+(Q'\cdot Q)^2U_4+Q^2(Q'\cdot Q)U_5,\\
Q'_\nu Q'_\mu\mathcal{M}^{c'_1,D_l^{*}\bar{D}_l^{*}}_{\nu\mu}&=U_3'=Q'^2U_1+Q'^2(Q'\cdot Q)U_2+Q'^2Q'^2U_3+Q'^2(Q'\cdot Q)U_4+(Q'\cdot Q)^2U_5,\\
Q_\nu Q'_\mu\mathcal{M}^{c'_1,D_l^{*}\bar{D}_l^{*}}_{\nu\mu}&=U_4'=(Q'\cdot Q)U_1+(Q'\cdot Q)^2U_2+Q'^2(Q'\cdot Q)U_3+Q^2Q'^2U_4+Q^2(Q'\cdot Q)U_5,\\
Q_\nu Q_\mu\mathcal{M}^{c'_1,D_l^{*}\bar{D}_l^{*}}_{\nu\mu}&=U_5'=Q^2U_1+Q^2(Q'\cdot Q)U_2+(Q'\cdot Q)^2U_3+Q^2(Q'\cdot Q)U_4+Q^2Q^2U_5,
\end{split}
\end{equation}
where $U'_i$ are numbers. Subsequently, we numerically calculate $U'_i$ and solve this set of equations. The values of $U_i$ can be obtained.

Then we can obtain the decay width with mass of meson-meson bound state for channel $J/\psi\omega$
\begin{equation}\label{decaywidthM0}
\begin{split}
\Gamma_1(M_0)=&\int d^3Qd^3Q'(2\pi)^4\delta^{(4)}(Q+Q'-P)\sum_{\varrho'=1}^3\sum_{\varrho=1}^3(2\pi)^{3}|T_{(c'_1;b)a}(M_0)|^2,
\end{split}
\end{equation}
where $P=(0,0,0,iM_0)$, $Q=(\mathbf{Q}(M_0),i\sqrt{\mathbf{Q}^2(M_0)+M^2_{J/\psi}})$, $Q'=(-\mathbf{Q}(M_0),i\sqrt{\mathbf{Q}^2(M_0)+M^2_\omega})$ and $\mathbf{Q}^2(M_0)=[M_0^2-(M_{J/\psi}+M_\omega)^2][M_0^2-(M_{J/\psi}-M_\omega)^2]/(4M_0^2)$. To calculate the decay width $\Gamma_1(M_0)$, we use the transverse condition $\varepsilon^{\varrho}(Q)\cdot Q=\varepsilon^{\varrho'}(Q')\cdot Q'=0$ and the completeness relation. It is necessary to emphasize that the decay width $\Gamma_1(M_0)$ expressed as Eq. (\ref{decaywidthM0}) is not the decay width of physical resonance.

\subsection{Decay channel $D^{+}D^{-}$ with respect to mass of bound state}\label{sec:DpDmbs}
\begin{figure}[!htb] \centering
\includegraphics[trim = 0mm 50mm 0mm 40mm,scale=1,width=10cm]{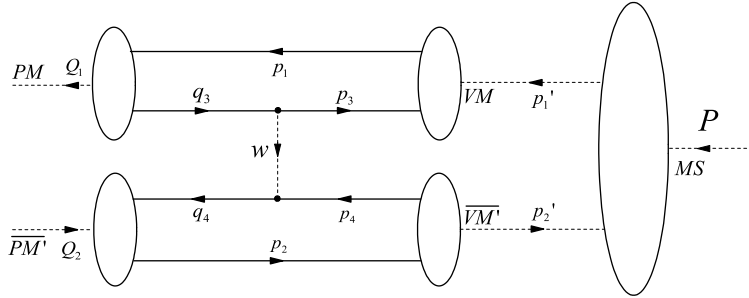}
\caption{\label{Fig4} Matrix element for decay channel $D^+D^-$. The momenta in the final state satisfy $Q_1+Q_2=P$. $w$ represents the momentum of the exchanged light meson.}
\end{figure}

Considering the lowest order term of the two-particle irreducible Green's function, we obtain the interaction between two heavy vector mesons derived from a light meson exchange. Applying Mandelstam's approach, we can obtain the $T$-matrix element with mass $M_0$ for channel $c'_2$, which can be represented graphically by Fig. \ref{Fig4}. In Fig. \ref{Fig4}, $PM$ and $\overline{PM'}$ represent pseudoscalar mesons $D^{+}$ and $D^{-}$, respectively; $Q_1$ and $Q_2$ represent the momenta of final particles, $Q_1^2=-M_{D^{+}}^2$, $Q_2^2=-M_{D^{-}}^2$ and in this section $Q_1+Q_2=P$.

\begin{figure}[!htb] \centering
\includegraphics[trim = 0mm 40mm 0mm 40mm,scale=1,width=10cm]{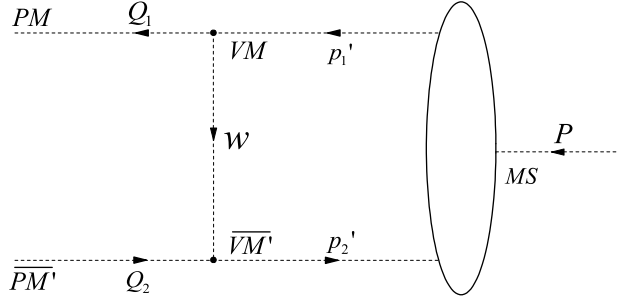}
\caption{\label{Fig5} Reduced matrix element for decay channel $D^{+}D^{-}$.}
\end{figure}

To simplify the computational process, we use the vertex function for the exchanged light meson, heavy pseudoscalar and vector mesons; and then Fig. \ref{Fig4} can be reduced to Fig. \ref{Fig5}. From the Lorentz-structure, we obtain the matrix elements of quark scalar density $J$ and quark current $J_\alpha$ between heavy pseudoscalar and vector mesons
\begin{subequations}\label{vertice1}
\begin{equation}
\begin{split}
\langle PM(Q_1)|J|VM^\vartheta(p_1') \rangle=&\frac{1}{\sqrt{2E_{D^+}(Q_1)}\sqrt{2E_{D^*_l}(p_1')}}[Q_1\cdot\varepsilon^\vartheta(p_1')]h^{(s)}_{4}(w^{2}),
\end{split}
\end{equation}
\begin{equation}
\begin{split}
\langle \overline{PM'}(Q_2)|J|\overline{VM'}^{\vartheta'}(-p_2')\rangle=&\frac{1}{2\sqrt{E_{D^-}(Q_2)E_{\bar D^*_l}(-p_2')}}[Q_2\cdot\varepsilon^{\vartheta'}(-p_2')]\bar{h}^{(s)}_{4}(w^{2}),
\end{split}
\end{equation}
\begin{equation}
\begin{split}
\langle &PM(Q_1)|J_{\alpha}|VM^\vartheta(p_1')\rangle=\frac{1}{2\sqrt{E_{D^+}(Q_1)E_{D^*_l}(p_1')}}\\
&\times\{h_{4}^{(\text{v})}(w^{2})\{[Q_1\cdot\varepsilon^\vartheta(p_1')](Q_1+p_1')_{\alpha}-[(Q_1+p_1')\cdot(Q_1-p_1')]\varepsilon_\alpha^\vartheta(p_1')\}\\
&-h^{(\text{v})}_{5}(w^{2})\{[Q_1\cdot\varepsilon^\vartheta(p_1')](Q_1-p_1')_{\alpha}-(Q_1-p_1')^2\varepsilon_\alpha^\vartheta(p_1')\}\},
\end{split}
\end{equation}
\begin{equation}
\begin{split}
\langle &\overline{PM'}(Q_2)|J_{\beta}|\overline{VM'}^{\vartheta'}(-p_2')\rangle=\frac{1}{2\sqrt{E_{D^-}(Q_2)E_{\bar D^*_l}(-p_2')}}\\
&\times\{\bar{h}_{4}^{(\text{v})}(w^{2})\{[Q_2\cdot\varepsilon^{\vartheta'}(-p_2')](Q_2-p_2')_{\beta}-[(Q_2-p_2')\cdot(Q_2+p_2')]\varepsilon^{\vartheta'}_{\beta}(-p_2')\}\\
&-\bar{h}^{(\text{v})}_{5}(w^{2})\{[Q_2\cdot\varepsilon^{\vartheta'}(-p_2')](Q_2+p_2')_{\beta}-(Q_2+p_2')^2\varepsilon^{\vartheta'}_{\beta}(-p_2')\}\},
\end{split}
\end{equation}
\end{subequations}
where $p'_1=(\mathbf{p},ip_{10}')$, $p'_2=(\mathbf{p},ip_{20}')$, $Q_1=(\mathbf{Q}_D,iQ_{10})$, $Q_2=(-\mathbf{Q}_D,iQ_{20})$, $w=p_1'-Q_1=p_2'+Q_2=p-(Q_1-Q_2)/2$ is the momentum of the light meson and $\mathbf{w}=\mathbf{p}-\mathbf{Q}_D$; $E_{D^\pm}(p)=\sqrt{\mathbf{p}^{2}+M_{D^\pm}^{2}}$, $h(w^2)$ and $\bar h(w^2)$ become vertex functions.

\begin{figure}[!htb] \centering
\includegraphics[trim = 0mm 30mm 0mm 20mm,scale=1,width=8cm]{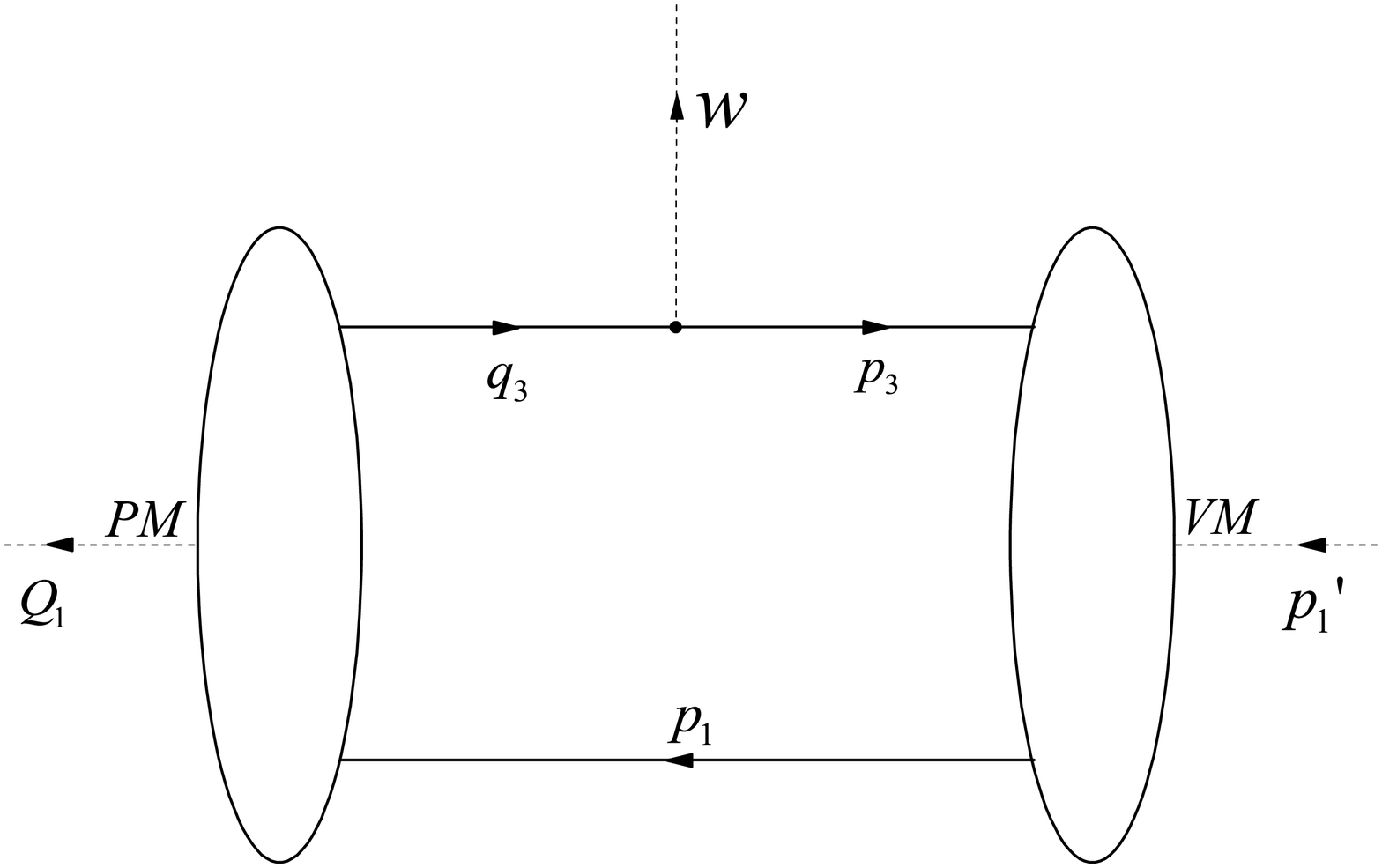}
\caption{\label{Fig6} Vertex function for the exchanged light meson, heavy pseudoscalar and vector mesons.}
\end{figure}

Now, we introduce the vertex function for the exchanged light meson, heavy pseudoscalar and vector mesons, shown as Fig. \ref{Fig6}. The charmed meson $D^{+}$ is composed of $c$-quark and $d$-antiquark. For heavy pseudoscalar mesons, the authors of Refs. \cite{BSE:Roberts1,BSE:Roberts3,BSE:Roberts4,BSE:Roberts5} also gave their BS amplitudes in Euclidean space:
\begin{equation}
\begin{split}
\Gamma^{P}(K,k)=\frac{1}{\mathcal{N}^{P}}i\gamma_5\varphi_{P}(k^2),
\end{split}
\end{equation}
where $K$ is the momentum of heavy meson, $k$ denotes the relative momentum between quark and antiquark in heavy meson, $\mathcal{N}^P$ is normalization, and $\varphi_{P}(k^{2})$ is scalar function fixed by providing fits to observables. Using the approach introduced in Sec. \ref{sec:Formfac}, we can obtain the heavy pseudoscalar meson wave function in instantaneous approximation
\begin{equation}
\begin{split}\label{wfP}
\Psi^{D^+}(\textbf{k})&=\int dk_4\frac{1}{4}\text{Tr}\bigg\{\gamma_5\frac{-1}{\gamma\cdot(k+K)-im_c}\frac{1}{\mathcal{N}^{D^+}}i\gamma_5\varphi_{D^+}(k^2)\frac{-1}{\gamma\cdot k-im_d}\bigg\}\\
&=\int dk_4\frac{-i}{N^{D^+}}\exp\bigg(\frac{-\textbf{k}^2-k_4^2}{\omega_{D}^2}\bigg)\frac{\textbf{k}^2+k_4^2+m_cm_d}
{(\textbf{k}^2+k_4^2+m_c^2)(\textbf{k}^2+k_4^2+m_d^2)},
\end{split}
\end{equation}
where $m_{c,d}$ are the constituent quark masses, $\varphi_{D^+}(k^{2})=\varphi_{D^-}(k^{2})=\exp(-k^{2}/\omega_{D}^{2})$ and $\omega_{D}$=1.50 GeV \cite{BSE:Roberts5}. Then we can apply the method given in Refs. \cite{mypaper3,mypaper4,mypaper5} to obtain the explicit forms of the vertex function for heavy pseudoscalar meson $D$ and vector meson $D_l^{*}$ coupling to scalar meson ($\sigma$)
\begin{equation}
\begin{split}
-\frac{h^{(s)}_{4}(w^{2})}{2E_1}=&\frac{\bar{h}_{4}^{(s)}(w^{2})}{2E_2}=F_4(\textbf{w}^2),\\
F_{4}(\textbf{w}^{2})=&\frac{1}{\sqrt{M_{D^{*+}}^2-M_{D^+}^2}}\int
\frac{d^{3}k}{(2\pi)^{3}}\bar{\Psi}^{D^+}\bigg(\textbf{k}+\frac{E_c(k)}{E_{D^+}}\textbf{w}\bigg)\sqrt{\frac{E_d(k)+m_d}{E_d(k+w)+m_d}}\\
&\times\left\{\frac{E_d(k+w)-E_d(k)+2m_d}{2\sqrt{E_d(k+w)E_d(k)}}-\frac{\textbf{k}\cdot\textbf{w}}{2\sqrt{E_d(k+w)E_d(k)}[E_d(k)+m_d]}\right\}\Psi^{D^{*+}}(\textbf{k}),
\end{split}
\end{equation}
and to vector meson ($\rho$, $V_1$ and $V_8$)
\begin{equation}
\begin{split}
h_{4}^{(\text{v})}(w^{2})=&h_{5}^{(\text{v})}(w^{2})=\bar{h}_{4}^{(\text{v})}(w^{2})=\bar{h}_{5}^{(\text{v})}(w^{2})=F_5({\textbf{w}^{2}}),\\
F_5(\textbf{w}^{2})=&\frac{1}{\sqrt{M_{D_l^*}^2-M_{D^+}^2}}\frac{2\sqrt{E_{D_l^*}E_{D^+}}}{E_{D_l^*}+E_{D^+}}\int \frac{d^{3}k}{(2\pi)^{3}}\bar{\Psi}^{D^+}\bigg(\textbf{k}+\frac{E_c(k)}{E_{D^+}}\textbf{w}\bigg)\sqrt{\frac{E_l(k)+m_l}{E_d(k+w)+m_d}}\\
&\times\left\{\frac{E_d(k+w)+E_l(k)}{2\sqrt{E_d(k+w)E_l(k)}}+\frac{\textbf{k}\cdot\textbf{w}}{2\sqrt{E_d(k+w)E_l(k)}[E_l(k)+m_l]}\right\}\Psi^{D_l^*}({\textbf{k}}),
\end{split}
\end{equation}
where $E_{d}(p)=\sqrt{\mathbf{p}^{2}+m_{d}^{2}}$, $\Psi^{D^+}$ and $\Psi^{D_l^*}$ are the wave functions of heavy pseudoscalar and vector mesons expressed as Eqs. (\ref{wfP}) and (\ref{wfV}), respectively.

Taking away the external lines including normalizations and polarization vectors $\varepsilon^\vartheta_{\lambda}(p_1')$, $\varepsilon^{\vartheta'}_{\tau}(-p_2')$ in Eq. (\ref{vertice1}), we obtain the interaction from one light meson ($\sigma$, $\rho^0$, $V_1$ and $V_8$) exchange
\begin{equation}\label{kernel2}
\begin{split}
\mathcal{V}_{\lambda\tau}(Q_1,Q_2,\textbf{p})
=&-2E_1F_4(\textbf{w}^2)\frac{-ig_\sigma^2}{w^2+M_\sigma^2}2E_2F_4(\textbf{w}^2)Q_{1\lambda}Q_{2\tau}\\
&-4F_5(\textbf{w}^2)\left(\frac{-ig_\rho^2}{w^2+M_\rho^2}+\frac{-ig_1^2}{w^2+M_\omega^2}+\frac{-ig_8^2}{w^2+M_\phi^2}\right)F_5(\textbf{w}^2)(p_1'\cdot p_2')Q_{1\lambda}Q_{2\tau},\\
\end{split}
\end{equation}
where $E_1=E_2=M_0/2$, and $w=(\mathbf{w},0)$. The interaction from one-$\rho^{\pm}$ exchange becomes
\begin{equation}\label{kernel2'}
\begin{split}
\mathcal{V}'_{\lambda\tau}(Q_1,Q_2,\textbf{p})
=-4F_5(\textbf{w}^2)\frac{-i2g_\rho^2}{w^2+M_\rho^2}F_5(\textbf{w}^2)(p_1'\cdot p_2')Q_{1\lambda}Q_{2\tau}.\\
\end{split}
\end{equation}
These momenta in Fig. \ref{Fig5} become
\begin{equation}\label{momenta2}
\begin{split}
w=(\mathbf{p}-\mathbf{Q}_{D}(M_0),0),~Q_1+Q_2=P,~p_1'=p+P/2,~p_2'=p-P/2,
\end{split}
\end{equation}
where $P=(0,0,0,iM_0)$, $Q_1=(\mathbf{Q}_{D}(M_0),iM_0/2)$, $Q_2=(-\mathbf{Q}_{D}(M_0),iM_0/2)$ and $\mathbf{Q}_{D}^2(M_0)=[M_0^2-(M_{D^+}+M_{D^-})^2]/4$. For decay channel $D^{+}D^{-}$, we obtain the total matrix element
\begin{equation}\label{Smatrele2}
\begin{split}
-iR_{(c'_2;b)a}(M_0)=\langle Q_1,Q_2~\text{out}|P~\text{in}\rangle=-i(2\pi)^4\delta^{(4)}(Q_1+Q_2-P)T_{(c'_2;b)a}(M_0),
\end{split}
\end{equation}
where $T_{(c'_2;b)a}(M_0)$ is the $T$-matrix element with mass $M_0$ for channel $c'_2$. From Fig. \ref{Fig5}, we obtain
\begin{equation}\label{Tmatrele2}
\begin{split}
T_{(c'_2;b)a}(M_0)=\frac{-i}{(2\pi)^{9/2}\sqrt{2E_{D^+}(Q_1)}\sqrt{2E_{D^-}(Q_2)}\sqrt{2E(P)}}\bigg(\frac{1}{\sqrt{2}}\mathcal{M}^{c'_2}+\frac{1}{\sqrt{2}}\mathcal{M}'^{c'_2}\bigg),
\end{split}
\end{equation}
where
\begin{subequations}\label{MEcurr2}
\begin{equation}
\begin{split}
\mathcal{M}^{c'_2}=&\int\frac{d^4p}{(2\pi)^4}\mathcal{V}_{\lambda\tau}(Q_1,Q_2,\mathbf{p})\chi_{\lambda\tau}^{0^+}(P,p),
\end{split}
\end{equation}
\begin{equation}
\begin{split}
\mathcal{M}'^{c'_2}=&\int\frac{d^4p}{(2\pi)^4}\mathcal{V}'_{\lambda\tau}(Q_1,Q_2,\mathbf{p})\chi_{\lambda\tau}^{0^+}(P,p).
\end{split}
\end{equation}
\end{subequations}
Here $\chi_{\lambda\tau}^{0^+}(P,p)$ is expressed as Eq. (\ref{bswf0+1}). The $p$ integral is also computed in instantaneous approximation. Then the decay width with mass of meson-meson bound state for channel $D^{+}D^{-}$ becomes
\begin{equation}\label{decaywidth2M0}
\begin{split}
\Gamma_2(M_0)=&\int d^3Q_1d^3Q_2(2\pi)^4\delta^{(4)}(Q_1+Q_2-P)(2\pi)^{3}|T_{(c'_2;b)a}(M_0)|^2.
\end{split}
\end{equation}
The decay width $\Gamma_2(M_0)$ also is not the width of physical resonance.

\subsection{Decay channel $D^{0}\bar D^{0}$ with respect to mass of bound state}\label{sec:D0D0BARbs}
Since the $\chi_{c0}(3915)$ state is a isoscalar, there should exist the neutral channel $D^{0}\bar D^{0}$. In Figs. \ref{Fig4}, \ref{Fig5} and \ref{Fig6}, $PM$ and $\overline{PM'}$ represent pseudoscalar mesons $D^{0}$ and $\bar D^{0}$, respectively. Following the same procedure as for charged channel $D^{+}D^{-}$, we can obtain the $T$-matrix element $T_{(c'_3;b)a}(M_0)$ and the decay width $\Gamma_3(M_0)$ with mass $M_0$ for neutral channel $c'_3$. The decay width $\Gamma_3(M_0)$ should not be the width of physical resonance.

\section{the developed Bethe-Salpeter theory}\label{sec:tetotalH}
Secs. \ref{sec:GBSwfmm} and \ref{sec:Sme} give the traditional technique to deal with molecular state in present particle physics. These masses of meson-meson bound states were regarded as masses of resonances \cite{ms:Swanson,ms:Torn,liu,mypaper4,Msigma2} and used to calculate decay widths of resonances \cite{mypaper6,mypaper7}, which should not be impeccable. To deal with resonance in the framework of relativistic quantum field theory, we considered the time evolution of molecular state as determined by the total Hamiltonian and provided the developed Bethe-Salpeter theory in Ref. \cite{mypaper8}.

Because the time evolution of molecular state is determined by the total Hamiltonian, exotic meson resonance should be considered as an unstable meson-meson molecular state. According to the developed Bethe-Salpeter theory for dealing with resonance \cite{mypaper8}, this unstable state has been prepared to decay at given time, and the prepared state can be regarded as a bound state with ground-state energy. Solving BS equation for arbitrary meson-meson bound state, one can obtain the mass $M_0$ and BS wave function $\chi_P(x'_1,x'_2)$ for this bound state with momentum $P=(\mathbf{P},i\sqrt{\mathbf{P}^2+M_0^2})$. Setting $t_1=0$ and $t_2=0$ in the ground-state BS wave function, we obtain a description for the prepared state (ps)
\begin{equation}
\begin{split}\label{BSWFT0}
\mathscr{X}^{\text{ps}}_a=\chi_{P}(\mathbf{x}_1',t_1=0,\mathbf{x}_2',t_2=0)=\frac{1}{(2\pi)^{3/2}}\frac{1}{\sqrt{2E(P)}}e^{i\mathbf{P}\cdot\mathbf{X}}\chi_P(\mathbf{X}').
\end{split}
\end{equation}

Now it is necessary to consider the total Hamiltonian
\begin{equation}
\begin{split}
H=K_I+V_I,
\end{split}
\end{equation}
where $K_I$ represents the interaction responsible for the formation of stationary bound state and $V_I$ stands for the interaction responsible for the decay of resonance. Then the time evolution of this system determined by the total Hamiltonian $H$ has the explicit form
\begin{equation}
\begin{split}\label{timeevo}
\mathscr{X}(t)=e^{-iHt}\mathscr{X}^{\text{ps}}_{a}=\frac{1}{2\pi i}\int_{C_2}d\epsilon e^{-i\epsilon t}\frac{1}{\epsilon-H}\mathscr{X}^{\text{ps}}_{a},
\end{split}
\end{equation}
where $(\epsilon-H)^{-1}$ is the Green's function and the contour $C_2$ runs from $ic_r+\infty$ to $ic_r-\infty$ in energy-plane. The positive constant $c_r$ is sufficiently large that no singularity of $(\epsilon-H)^{-1}$ lies above $C_2$. The time-dependent wave function $\mathscr{X}(t)$ provides a complete description of the system for $t>0$. Since $H\neq K_I$, this system should not remain in the prepared state $\mathscr{X}^{\text{ps}}_a$. Then at arbitrary time $t$ the probability amplitude of finding the system in the state $\mathscr{X}^{\text{ps}}_a$ is
\begin{equation}
\begin{split}
\mathscr{A}_a=(\mathscr{X}^{\text{ps}}_{a},\mathscr{X}(t))=\frac{1}{2\pi i}\int_{C_2}d\epsilon\frac{e^{-i\epsilon t}}{\epsilon-M_0-(2\pi)^3T_{aa}(\epsilon)}.
\end{split}
\end{equation}
In field theory the operator $T(\epsilon)$ is just the scattering matrix with energy $\epsilon$, and $T_{aa}(\epsilon)$ is the $T$-matrix element between two bound states, which is defined as
\begin{equation}
\begin{split}
\langle a~\text{out}|a~\text{in}\rangle=\langle a~\text{in}|a~\text{in}\rangle-i(2\pi)^4\delta^{(4)}(P-P)T_{aa}(\epsilon).
\end{split}
\end{equation}

Because of the analyticity of $T_{aa}(\epsilon)$, we define
\begin{equation}
\begin{split}\label{Taepsilon}
T_{aa}(\epsilon)=\mathbb{D}(\epsilon)-i\mathbb{I}(\epsilon),
\end{split}
\end{equation}
where $\epsilon$ approaches the real axis from above, $\mathbb{D}$ and $\mathbb{I}$ are the real and imaginary parts, respectively. In experiments, many exotic particles are narrow states and their decay widths are very small compared with their energy levels, i.e., $(2\pi)^3\mathbb{I}(M_0)\ll M_0$. This situation is ordinarily interpreted as implying that both $(2\pi)^3|\mathbb{D}(\epsilon)|$ and $(2\pi)^3\mathbb{I}(\epsilon)$ are also very small quantities, as compared to $M_0$. Therefore, we can expect that $[\epsilon-M_0-(2\pi)^3T_{aa}(\epsilon)]^{-1}$ has a pole on the second Riemann sheet
\begin{equation}
\begin{split}\label{pole}
\epsilon_{\text{pole}}\cong M_0+(2\pi)^3[\mathbb{D}(M_0)-i\mathbb{I}(M_0)]=M-i\frac{\Gamma(M_0)}{2},
\end{split}
\end{equation}
where $\Delta M=(2\pi)^3\mathbb{D}(M_0)$ is the correction for energy level of resonance and $M=M_0+(2\pi)^3\mathbb{D}(M_0)$ is the physical mass for resonance. This pole at $\epsilon_{\text{pole}}$ describes the resonance. The mass $M_0$ of two-body bound state is obtained by solving homogeneous BS equation, which should not be the mass of physical resonance. $\Gamma(M_0)$ with mass $M_0$ also should not be the width of physical resonance, which should depend on its physical mass $M$. We will minutely show the computational process of $T$-matrix element between two bound states $T_{aa}(\epsilon)$ in the next section.

\section{$T$-matrix element $T_{aa}(\epsilon)$}\label{sec:ac}
When there is only one decay channel, we can use the unitarity of $T_{aa}(\epsilon)$ to obtain \cite{GreenFun}
\begin{equation}
\begin{split}\label{Taepsilon1}
2\mathbb{I}(\epsilon)=\sum_b(2\pi)^4\delta^{(3)}(\mathbf{P}_b-\mathbf{P})\delta(E_b-\epsilon)|T_{ba}(\epsilon)|^2,
\end{split}
\end{equation}
where $P_b=(\mathbf{P}_b,iE_b)$ is the total energy-momentum vector of all particles in the final state and the $T$-matrix element $T_{ba}(\epsilon)$ is defined as $\langle b~\text{out}|a~\text{in}\rangle=-i(2\pi)^4\delta^{(3)}(\mathbf{P}_b-\mathbf{P})\delta(E_b-\epsilon)T_{ba}(\epsilon)$. The delta-function in Eq. (\ref{Taepsilon1}) means that the energy $\epsilon$ in scattering matrix is equal to the total energy $E_b$ of the final state, and $\sum_b$ represents summing over momenta and spins of all particles in the final state. For $E_b=\epsilon$, we also denote the total energy of the final state by $\epsilon$ and $\mathbb{I}(\epsilon)$ becomes a function of the final state energy. Using dispersion relation for the function $T_{aa}(\epsilon)$, we obtain
\begin{equation}
\begin{split}\label{disrel}
\mathbb{D}(\epsilon)=-\frac{\mathcal{P}}{\pi}\int_{\epsilon_M}^\infty \frac{\mathbb{I}(\epsilon')}{\epsilon'-\epsilon}d\epsilon',
\end{split}
\end{equation}
where the symbol $\mathcal{P}$ means that this integral is a principal value integral and the variable of integration is the total energy $\epsilon'$ of the final state. To calculate the real part, we need calculate the function $\mathbb{I}(\epsilon')$ of value of the final state energy $\epsilon'$, which is an arbitrary real number over the real interval $\epsilon_M<\epsilon'<\infty$. As usual the momentum of initial bound state $a$ is set as $P=(0,0,0,iM_0)$ in the rest frame and $\epsilon_M$ denotes the sum of all particle masses in the final state. We suppose that the final state $b$ may contain $n$ composite particles and $n'$ elementary particles in decay channel $c'$. From Eq. (\ref{Taepsilon1}), we have
\begin{equation}
\begin{split}\label{Iepsilon'}
\mathbb{I}(\epsilon')=&\frac{1}{2}\int d^3Q'_1\cdots d^3Q'_{n'}d^3Q_1\cdots d^3Q_n(2\pi)^4\delta^{(4)}(Q'_1+\cdots+Q_n-P^{\epsilon'})\sum_{\text{spins}}|T_{(c';b)a}(\epsilon')|^2,
\end{split}
\end{equation}
where $Q'_1\cdots Q'_{n'}$ and $Q_1\cdots Q_n$ are the momenta of final elementary and composite particles, respectively; $P^{\epsilon'}=(0,0,0,i\epsilon')$, $T_{(c';b)a}(\epsilon')$ is the $T$-matrix element with respect to $\epsilon'$, and $\sum_{\text{spins}}$ represents summing over spins of all particles in the final state. In Eq. (\ref{Iepsilon'}) the energy in scattering matrix is equal to the total energy $\epsilon'$ of the final state $b$, which is an arbitrary real number over the real interval $\epsilon_M<\epsilon'<\infty$. The mass $M_0$ and BS amplitude of initial bound state $a$ have been specified and the value of the initial state energy in the rest frame is a specified value $M_0$. From Eq. (\ref{Iepsilon'}), we have $\mathbb{I}(\epsilon')>0$ for $\epsilon'>\epsilon_M$ and $\mathbb{I}(\epsilon')=0$ for $\epsilon'\leqslant\epsilon_M$, which is the reason that the integration in dispersion relation (\ref{disrel}) ranges from $\epsilon_M$ to $+\infty$.

If there are several decay channels, we should write instead
\begin{equation}
\begin{split}
\mathbb{I}(\epsilon')=&\frac{1}{2}\sum_{c'}\int d^3Q'_1\cdots d^3Q'_{n'}d^3Q_1\cdots d^3Q_n(2\pi)^4\delta^{(4)}(Q'_1+\cdots+Q_n-P^{\epsilon'})\sum_{\text{spins}}|T_{(c';b)a}(\epsilon')|^2,
\end{split}
\end{equation}
where $\sum_{c'}$ represents summing over all open and closed channels. Because the total energy $\epsilon'$ of the final state extends from $\epsilon_M$ to $+\infty$, we may obtain several closed channels derived from the interaction Lagrangian. Assuming that resonance $\chi_{c0}(3915)$ is a mixed state of two components $D^{*0}\bar{D}^{*0}$ and $D^{*+}D^{*-}$, we obtain one closed channel $D^{*}\bar D^{*}$ derived from the interaction Lagrangian (\ref{Lag}), denoted as $c'_4$. Since bound state lies below the threshold, i.e., $M_0<M_{D^*}+M_{\bar D^*}$, the closed channel $c'_4$ can not occur inside the physical world.

\subsection{Channel $J/\psi\omega$ with respect to arbitrary value of the final state energy}\label{sec:Jpsioe'}
\begin{figure}[!htb] \centering
\includegraphics[trim = 0mm 30mm 0mm 30mm,scale=1,width=10cm]{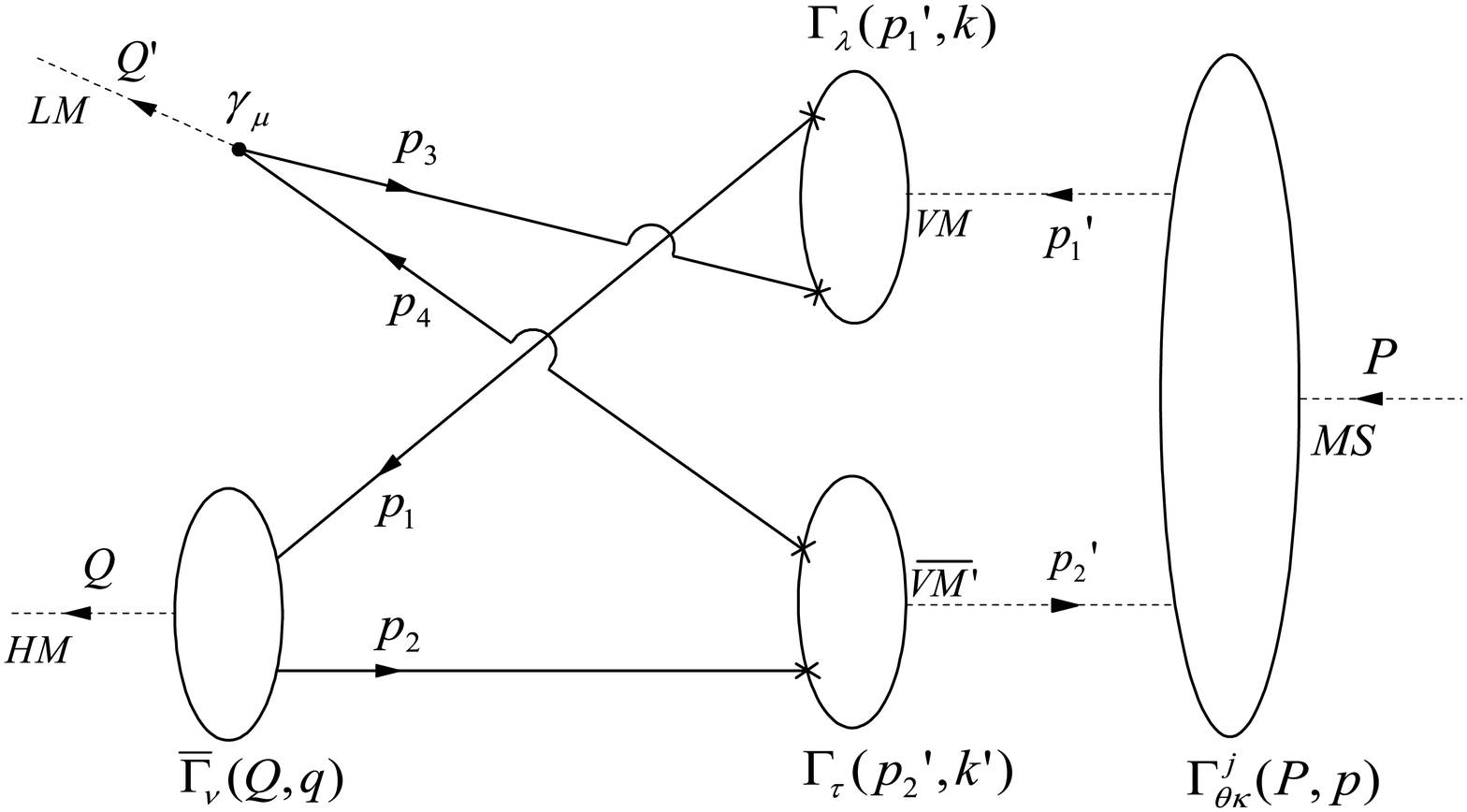}
\caption{\label{Fig7} Matrix element with respect to $\epsilon'$ for channel $J/\psi\omega$. The momenta in the final state satisfy $Q+Q'=P^{\epsilon'}$ and the momentum of the initial state is $P$. The final state energy extends from $\epsilon_{M}$ to $+\infty$ while the initial state energy is specified, and the crosses mean that the momenta of quark propagators depend on the final state energy $\epsilon'$.}
\end{figure}

From Eq. (\ref{Smatrele}), we obtain the total matrix element between the final state $\langle J/\psi(Q),\omega(Q')~\text{out}|$ and the specified initial four-quark state $|P~\text{in}\rangle$
\begin{equation}
\begin{split}
-iR_{(c'_1;b)a}(\epsilon')=\langle Q,Q'~\text{out}|P~\text{in}\rangle=-i(2\pi)^4\delta^{(4)}(Q+Q'-P^{\epsilon'})T_{(c'_1;b)a}(\epsilon'),
\end{split}
\end{equation}
where the total energy $\epsilon'$ of the final state extends from $\epsilon_{c'_1,M}$ to $+\infty$, i.e., $\epsilon_{c'_1,M}<\epsilon'<\infty$ and $\epsilon_{c'_1,M}=M_{J/\psi}+M_\omega$. $T_{(c'_1;b)a}(\epsilon')$ is the bound state matrix element with respect to $\epsilon'$ for channel $c'_1$, shown as Fig. \ref{Fig7}. It is necessary to emphasize that the energy in the two-particle irreducible Green's function is equal to the final state energy $\epsilon'$ while the mass $M_0$ and BS amplitude of initial bound state is specified. We have introduced \emph{extended Feynman diagram} in Ref. \cite{mypaper8} to represent arbitrary value of the final state energy. In Fig. \ref{Fig7}, the quark momenta in left-hand side of crosses depend on the final state energy and the momenta in right-hand side depend on the initial state energy, i.e., $p_1-p_2-p_3+p_4=Q+Q'=P^{\epsilon'}$ and $p_1'-p_2'=P$. When $\epsilon'=M_0$, the crosses in Fig. \ref{Fig7} disappear and then Fig. \ref{Fig7} becomes Fig. \ref{Fig3}; $T_{(c'_1;b)a}(\epsilon'=M_0)$ is the $T$-matrix element with mass $M_0$ for channel $c'_1$ expressed as Eq. (\ref{Tmatrele}). Though the $T$-matrix element $T_{(c'_1;b)a}(\epsilon')$ has the same form expressed as Eq. (\ref{Tmatrele}), these momenta should become
\begin{equation}\label{momenta}
\begin{split}
&p_1=(Q+Q')/2+p+k,~~p_2=(Q+Q')/2-Q+p+k,~~p_3=k,~~p_4=Q'+k,\\
&q=Q'/2+p+k,~k'=Q'(M_0)+k,~~p_1'=p+P/2,~~p_2'=p-P/2,~~Q+Q'=P^{\epsilon'},
\end{split}
\end{equation}
where $P=(0,0,0,iM_0)$, $P^{\epsilon'}=(0,0,0,i\epsilon')$, $Q'(M_0)=(-\mathbf{Q}(M_0),i\sqrt{\mathbf{Q}^2(M_0)+M_\omega^2})$, $Q=(\mathbf{Q}(\epsilon'),i\sqrt{\mathbf{Q}^2(\epsilon')+M_{J/\psi}^2})$, $Q'=(-\mathbf{Q}(\epsilon'),i\sqrt{\mathbf{Q}^2(\epsilon')+M_\omega^2})$ and $\mathbf{Q}^2(\epsilon')=[\epsilon'^2-(M_{J/\psi}+M_\omega)^2][\epsilon'^2-(M_{J/\psi}-M_\omega)^2]/(4\epsilon'^2)$. The initial state is considered as a four-quark state, so the specified GBS amplitude of initial state should be
\begin{equation}
\begin{split}
\Gamma^{D_l^{*}}_\lambda(p_1',k)\chi^{0^+}_{\lambda\tau}(P,p)\Gamma^{\bar{D}_l^{*}}_\tau(p_2',k'),
\end{split}
\end{equation}
where $k'$ depends on $P$. Then we obtain the function $\mathbb{I}_1(\epsilon')$ for channel $J/\psi\omega$
\begin{equation}\label{Iepsilon'1}
\begin{split}
\mathbb{I}_1(\epsilon')&=\frac{1}{2}\int d^3Qd^3Q'(2\pi)^4\delta^{(4)}(Q+Q'-P^{\epsilon'})\sum_{\varrho'=1}^3\sum_{\varrho=1}^3|T_{(c'_1;b)a}(\epsilon')|^2.
\end{split}
\end{equation}

\subsection{Channel $D^{+}D^{-}$ with respect to arbitrary value of the final state energy}\label{sec:DpDme'}
\begin{figure}[!htb] \centering
\includegraphics[trim = 0mm 50mm 0mm 40mm,scale=1,width=10cm]{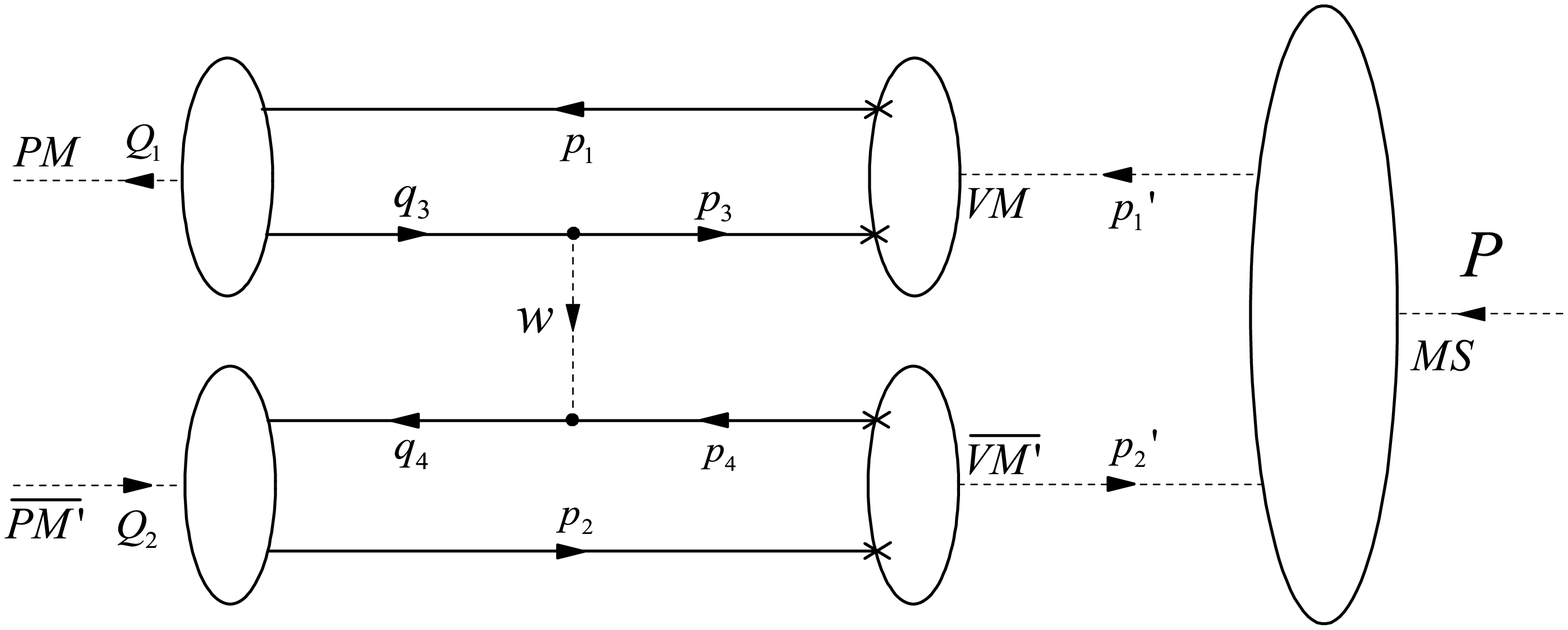}
\caption{\label{Fig8} Matrix element with respect to $\epsilon'$ for channel $D^+D^-$. The momenta in the final state satisfy $Q_1+Q_2=P^{\epsilon'}$ and the momentum of the initial state is $P$. $w$ represents the momentum of the exchanged light meson. The crosses mean that the momenta of quark propagators and the momentum $w$ of the exchanged light meson depend on the final state energy $\epsilon'$.}
\end{figure}

The $T$-matrix element with respect to $\epsilon'$ for channel $c'_2$ can be represented graphically by Fig. \ref{Fig8}. The total energy $\epsilon'$ of the final state extends from $\epsilon_{c'_2,M}$ to $+\infty$, i.e., $\epsilon_{c'_2,M}<\epsilon'<\infty$ and $\epsilon_{c'_2,M}=M_{D^+}+M_{D^-}$. In Fig. \ref{Fig8}, the crosses mean that the momenta of quark propagators and the momentum $w$ of the exchanged light meson depend on $Q_1$ and $Q_2$, i.e., $p_1-p_2-p_3+p_4=p_1-p_2-q_3+q_4=Q_1+Q_2=P^{\epsilon'}$ and $p_1'-p_2'=P$.

\begin{figure}[!htb] \centering
\includegraphics[trim = 0mm 30mm 0mm 20mm,scale=1,width=8cm]{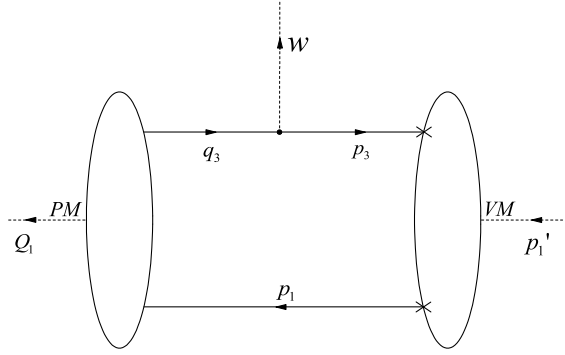}
\caption{\label{Fig9} Vertex function for the exchanged light meson, heavy pseudoscalar and vector mesons with respect to $\epsilon'$. $Q_1$ depends on $P^{\epsilon'}$ and $p_1'$ depends on $P$. The crosses mean that the momenta of quark propagators and the momentum $w$ of the exchanged light meson depend on the final state energy $\epsilon'$.}
\end{figure}

\begin{figure}[!htb] \centering
\includegraphics[trim = 0mm 40mm 0mm 40mm,scale=1,width=10cm]{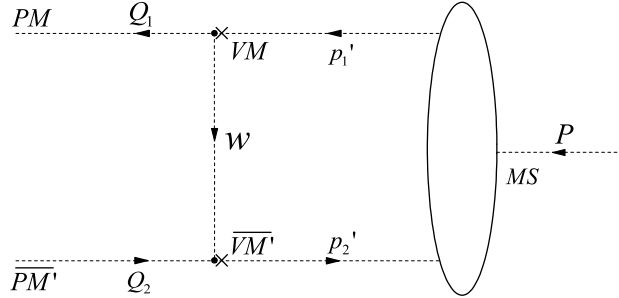}
\caption{\label{Fig10} Reduced matrix element with respect to $\epsilon'$ for channel $D^{+}D^{-}$. The crosses mean that the momentum $w$ of the exchanged light meson depends on $Q_1$ and $Q_2$.}
\end{figure}

We still use the vertex function to calculate the $T$-matrix element with respect to $\epsilon'$ for channel $c'_2$. However, different from the ordinary vertex function, we should introduce the vertex function with respect to $\epsilon'$, which is shown as Fig. \ref{Fig9}. In Fig. \ref{Fig9}, $Q_1$ depends on $P^{\epsilon'}$, $p_1'$ depends on $P$ and the crosses mean that the momenta of quark propagators and the momentum $w$ of the exchanged light meson depend on the final state energy $\epsilon'$. Using the approach introduced in Sec. \ref{sec:DpDmbs}, we can obtain the explicit forms for the vertex functions with respect to $\epsilon'$, and then Fig. \ref{Fig8} can be reduced to Fig. \ref{Fig10}. In Fig. \ref{Fig10}, we have $Q_1+Q_2=P^{\epsilon'}$, $p_1'-p_2'=P$ and the crosses lie on the right-hand side of light meson propagator, which implies that the momentum $w$ of the exchanged light meson depends on $Q_1$ and $Q_2$.

From Eq. (\ref{Smatrele2}), we obtain the total matrix element between the final state $\langle D^+(Q_1),D^-(Q_2)~\text{out}|$ and the mixed state of two pure bound states $D^{*0}\bar{D}^{*0}$ and $D^{*+}D^{*-}$
\begin{equation}
\begin{split}
-iR_{(c'_2;b)a}(\epsilon')=\langle Q_1,Q_2~\text{out}|P~\text{in}\rangle=-i(2\pi)^4\delta^{(4)}(Q_1+Q_2-P^{\epsilon'})T_{(c'_2;b)a}(\epsilon'),
\end{split}
\end{equation}
where $T_{(c'_2;b)a}(\epsilon')$ is the bound state matrix element with respect to $\epsilon'$ for channel $c'_2$, shown as Fig. \ref{Fig10}. When $\epsilon'=M_0$, the crosses in Figs. \ref{Fig8}, \ref{Fig9} and \ref{Fig10}  disappear and then these three extended Feynman diagrams become Figs. \ref{Fig4}, \ref{Fig6} and \ref{Fig5}, respectively; $T_{(c'_2;b)a}(\epsilon'=M_0)$ is the $T$-matrix element with mass $M_0$ for channel $c'_2$ expressed as Eq. (\ref{Tmatrele2}). Though the $T$-matrix element $T_{(c'_2;b)a}(\epsilon')$ has the same form expressed as Eq. (\ref{Tmatrele2}), these momenta should become
\begin{equation}\label{momenta2}
\begin{split}
&w=(\mathbf{p}-\mathbf{Q}_{D}(\epsilon'),0),~Q_1+Q_2=P^{\epsilon'},~p_1'=p+P/2,~p_2'=p-P/2,
\end{split}
\end{equation}
where $P=(0,0,0,iM_0)$, $P^{\epsilon'}=(0,0,0,i\epsilon')$, $Q_1=(\mathbf{Q}_{D}(\epsilon'),i\epsilon'/2)$, $Q_2=(-\mathbf{Q}_{D}(\epsilon'),i\epsilon'/2)$ and $\mathbf{Q}_{D}^2(\epsilon')=[\epsilon'^2-(M_{D^+}+M_{D^-})^2]/4$. The coefficients $E_1$ and $E_2$ in interaction $\mathcal{V}_{\lambda\tau}(Q_1,Q_2,\textbf{p})$ given by Eq. (\ref{kernel2}) should become $E_1(\epsilon')=E_2(\epsilon')=\sqrt{\epsilon'M_0}/2$. Then we obtain the function $\mathbb{I}_2(\epsilon')$ for channel $D^+D^-$
\begin{equation}\label{Iepsilon'2}
\begin{split}
\mathbb{I}_2(\epsilon')&=\frac{1}{2}\int d^3Q_1d^3Q_2(2\pi)^4\delta^{(4)}(Q_1+Q_2-P^{\epsilon'})|T_{(c'_2;b)a}(\epsilon')|^2.
\end{split}
\end{equation}

\subsection{Channel $D^{0}\bar D^{0}$ with respect to arbitrary value of the final state energy}\label{sec:D0D0BARe'}
In Figs. \ref{Fig8}, \ref{Fig9} and \ref{Fig10}, $PM$ and $\overline{PM'}$ represent pseudoscalar mesons $D^{0}$ and $\bar D^{0}$, respectively. Following the same procedure as for charged channel $D^{+}D^{-}$, we can obtain the $T$-matrix element $T_{(c'_3;b)a}(\epsilon')$ with respect to $\epsilon'$ and the function $\mathbb{I}_3(\epsilon')$ for neutral channel $c'_3$. Here, the total energy $\epsilon'$ of the final state extends from $\epsilon_{c'_3,M}$ to $+\infty$, i.e., $\epsilon_{c'_3,M}<\epsilon'<\infty$ and $\epsilon_{c'_3,M}=M_{D^0}+M_{\bar D^0}$.

\subsection{Closed channel $D^{*}\bar D^{*}$}\label{sec:DDBARe'}
\begin{figure}[!htb] \centering
\includegraphics[trim = 0mm 50mm 0mm 40mm,scale=1,width=10cm]{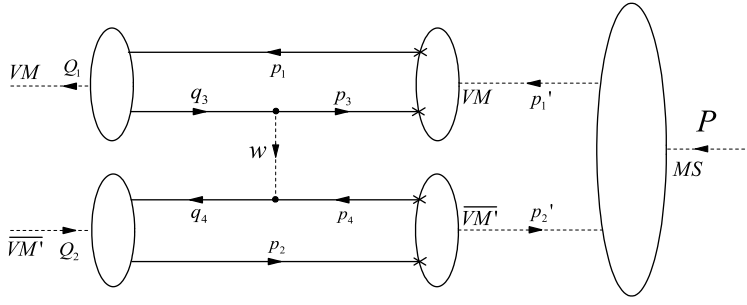}
\caption{\label{Fig11} Matrix element for closed channel $D_l^*\bar D_l^*$. The momenta in the final state satisfy $Q_1+Q_2=P^{\epsilon'}$ and the momentum of the initial state is $P$. $w$ represents the momentum of the exchanged light meson. The crosses mean that the momenta of quark propagators and the momentum $w$ of the exchanged light meson depend on the final state energy $\epsilon'$.}
\end{figure}

The final state $\langle D^*,\bar D^*~\text{out}|$ can be written as
\begin{equation}
\begin{split}
\langle D^*,\bar D^*~\text{out}|=\frac{1}{\sqrt2}\langle D^{*0},\bar D^{*0}~\text{out}|+\frac{1}{\sqrt2}\langle D^{*+},D^{*-}~\text{out}|.
\end{split}
\end{equation}
The total energy $\epsilon'$ of the final state extends from $\epsilon_{c'_4,M}$ to $+\infty$, i.e., $\epsilon_{c'_4,M}<\epsilon'<\infty$ and $\epsilon_{c'_4,M}=M_{D^*_l}+M_{\bar D^*_l}$. Considering the lowest order term of the two-particle irreducible Green's function, we can obtain the $T$-matrix element between the final state $\langle D^{*}_l,\bar D^{*}_l~\text{out}|$ and the initial four-quark state, which can be represented graphically by Fig. \ref{Fig11}. In Fig. \ref{Fig11}, $VM$ and $\overline{VM'}$ still represent $D_l^{*}$ and $\bar{D}_l^{*}$, respectively; $Q_1$ and $Q_2$ still represent the momenta of final particles, but $Q_1^2=-M_{D_l^{*}}^2$ and $Q_2^2=-M_{\bar D_l^{*}}^2$; the crosses mean that the momenta of quark propagators and the momentum $w$ of the exchanged light meson depend on $Q_1$ and $Q_2$, i.e., $p_1-p_2-p_3+p_4=p_1-p_2-q_3+q_4=Q_1+Q_2=P^{\epsilon'}$, and $p_1'-p_2'=P$.

\begin{figure}[!htb] \centering
\includegraphics[trim = 0mm 30mm 0mm 20mm,scale=1,width=8cm]{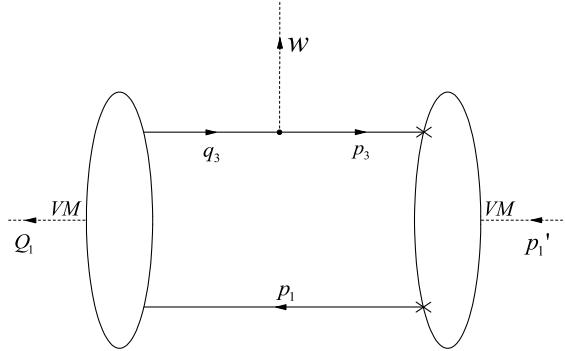}
\caption{\label{Fig12} Heavy meson form factor with respect to $\epsilon'$. $Q_1$ depends on $P^{\epsilon'}$ and $p_1'$ depends on $P$. The crosses mean that the momenta of quark propagators and the momentum $w$ of the exchanged light meson depend on the final state energy $\epsilon'$.}
\end{figure}

\begin{figure}[!htb] \centering
\includegraphics[trim = 0mm 40mm 0mm 40mm,scale=1,width=10cm]{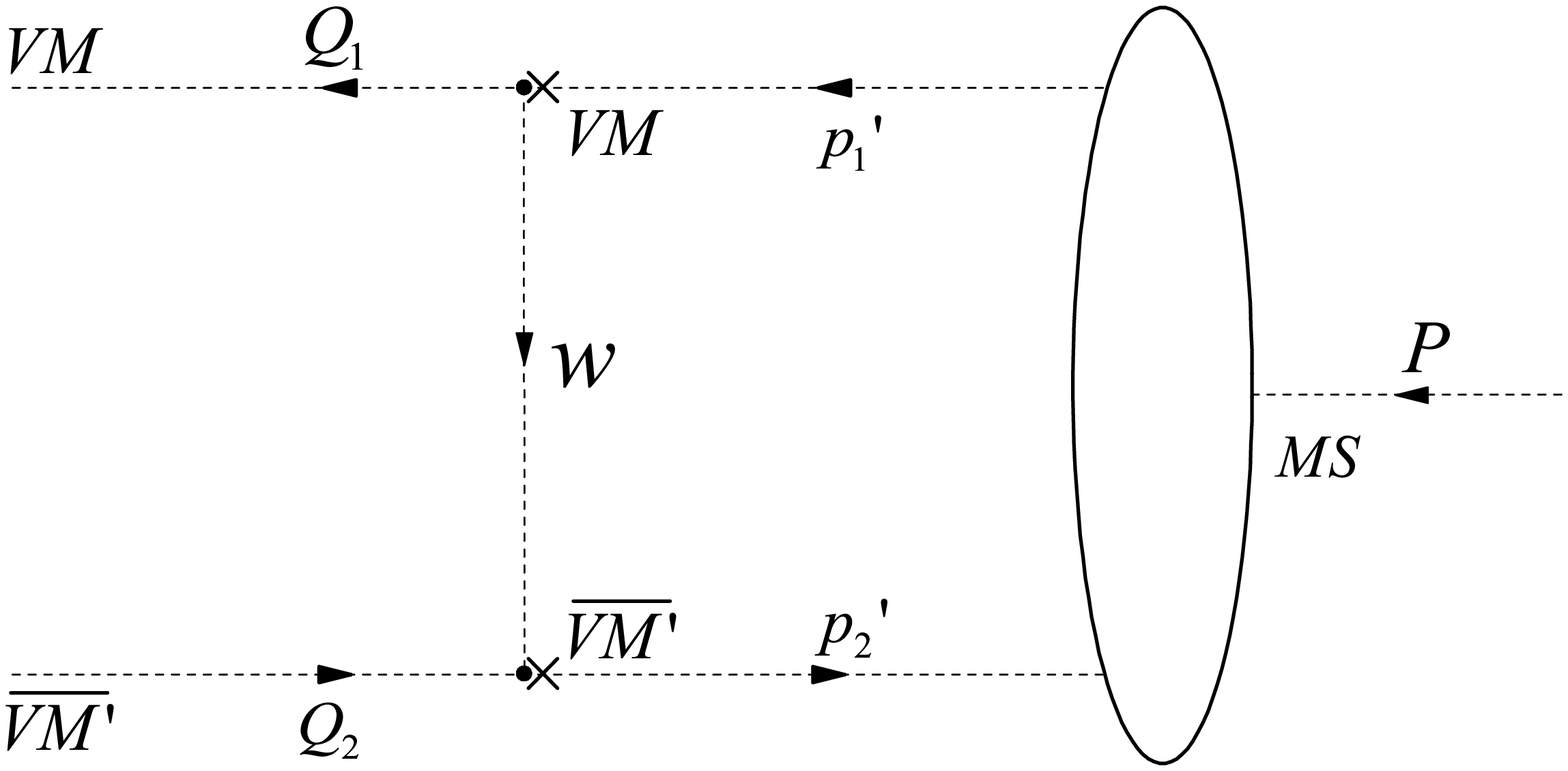}
\caption{\label{Fig13} Reduced matrix element for closed channel $D_l^{*}\bar D_l^{*}$. The crosses mean that the momentum $w$ of the exchanged light meson depends on $Q_1$ and $Q_2$.}
\end{figure}

To calculate the $T$-matrix element with respect to $\epsilon'$ for channel $c'_4$, we also introduce the form factor of heavy meson with respect to $\epsilon'$, which is shown as Fig. \ref{Fig12}. Using the approach introduced in Sec. \ref{sec:Formfac}, we can obtain the explicit forms for the heavy meson form factors $h(w^2)$ with respect to $\epsilon'$, and then Fig. \ref{Fig11} can be reduced to Fig. \ref{Fig13}. In Fig. \ref{Fig13}, we have $Q_1+Q_2=P^{\epsilon'}$, $p_1'-p_2'=P$, and the crosses lie on the right-hand side of light meson propagator, which implies that the momentum $w$ of the exchanged light meson depends on $Q_1$ and $Q_2$.

Using the Heisenberg picture, we obtain the total matrix element between the final state $\langle D^{*}_l(Q_1),\bar D^{*}_l(Q_2)~\text{out}|$ and the mixed state of two pure bound states $D^{*0}\bar{D}^{*0}$ and $D^{*+}D^{*-}$
\begin{equation}
\begin{split}
-iR_{(c'_4;b)a}(\epsilon')=\langle Q_1,Q_2~\text{out}|P~\text{in}\rangle=-i(2\pi)^4\delta^{(4)}(Q_1+Q_2-P^{\epsilon'})T_{(c'_4;b)a}(\epsilon').
\end{split}
\end{equation}
According to Mandelstam's approach, the $T$-matrix element becomes
\begin{equation}\label{Tmatrele3}
\begin{split}
T_{(c'_4;b)a}(\epsilon')=\frac{1}{2}\sum_{l=u,d}\frac{-i\varepsilon_\mu^{\varrho'}(Q_2)\varepsilon_\nu^{\varrho}(Q_1)}{(2\pi)^{9/2}\sqrt{2E_{D^*_l}(Q_1)}\sqrt{2E_{\bar D^*_l}(Q_2)}\sqrt{2E(P)}}(\mathcal{M}^{c'_4}_{\nu\mu}+\mathcal{M}'^{c'_4}_{\nu\mu}),
\end{split}
\end{equation}
where $\varepsilon_\nu^{\varrho=1,2,3}(Q_1)$ and $\varepsilon_\mu^{\varrho'=1,2,3}(Q_2)$ become the polarization vectors of $D^*_l$ and $\bar D^*_l$, respectively, and
\begin{subequations}\label{MEcurr3}
\begin{equation}
\begin{split}
\mathcal{M}^{c'_4}_{\nu\mu}=&\int \frac{d^4p}{(2\pi)^4}\mathcal{V}_{\nu\lambda,\tau\mu}(Q_1,Q_2,\mathbf{p})\chi_{\lambda\tau}^{0^+}(P,p),
\end{split}
\end{equation}
\begin{equation}
\begin{split}
\mathcal{M}'^{c'_4}_{\nu\mu}=&\int \frac{d^4p}{(2\pi)^4}\mathcal{V}'_{\nu\lambda,\tau\mu}(Q_1,Q_2,\mathbf{p})\chi_{\lambda\tau}^{0^+}(P,p).
\end{split}
\end{equation}
\end{subequations}
Here $\chi_{\lambda\tau}^{0^+}(P,p)$ is expressed as Eq. (\ref{bswf0+1}), $\mathcal{V}_{\nu\lambda,\tau\mu}(Q_1,Q_2,\mathbf{p})$ and $\mathcal{V}'_{\nu\lambda,\tau\mu}(Q_1,Q_2,\mathbf{p})$ represent the interactions derived from one light meson ($\sigma$, $\rho^0$, $V_1$ and $V_8$) exchange and one-$\rho^{\pm}$ exchange, respectively.

Now, we determine the interactions $\mathcal{V}_{\nu\lambda,\tau\mu}(Q_1,Q_2,\mathbf{p})$ and $\mathcal{V}'_{\nu\lambda,\tau\mu}(Q_1,Q_2,\mathbf{p})$. Structurally similar to Eq. (\ref{vertice}), we obtain the vertices of the heavy vector meson interaction with light meson
\begin{subequations}\label{meofqc}
\begin{equation}
\begin{split}
\langle VM^\varrho(Q_1)|J|VM^\vartheta(p_1') \rangle=&\frac{1}{2\sqrt{E_{D^*_l}(Q_1)E_{D^*_l}(p_1')}}[\varepsilon^\varrho(Q_1)\cdot\varepsilon^\vartheta(p_1')]h^{(\text{s})}_{1}(w^{2}),
\end{split}
\end{equation}
\begin{equation}
\begin{split}
\langle \overline{VM'}^{\varrho'}(Q_2)|J|\overline{VM'}^{\vartheta'}(-p_2')\rangle=&\frac{1}{2\sqrt{E_{\bar D^*_l}(Q_2)E_{\bar D^*_l}(-p_2')}}[\varepsilon^{\varrho'}(Q_2)\cdot\varepsilon^{\vartheta'}(-p_2')]\bar{h}^{(\text{s})}_{1}(w^{2}),
\end{split}
\end{equation}
\begin{equation}
\begin{split}
\langle VM^\varrho(Q_1)|J_{\alpha}|VM^\vartheta(p_1')\rangle=&\frac{1}{2\sqrt{E_{D^*_l}(Q_1)E_{D^*_l}(p_1')}}\{[\varepsilon^\varrho(Q_1)\cdot\varepsilon^\vartheta(p_1')]h_{1}^{(\text{v})}(w^{2})(Q_1+p_1')_{\alpha}\\
&-h^{(\text{v})}_{2}(w^{2})\{[\varepsilon^\varrho(Q_1)\cdot p_1']\varepsilon^\vartheta_{\alpha}(p_1')+[\varepsilon^\vartheta(p_1')\cdot Q_1]\varepsilon^\varrho_{\alpha}(Q_1)\}\},
\end{split}
\end{equation}
\begin{equation}
\begin{split}
\langle \overline{VM'}&^{\varrho'}(Q_2)|J_{\beta}|\overline{VM'}^{\vartheta'}(-p_2')\rangle\\
=&\frac{1}{2\sqrt{E_{\bar D^*_l}(Q_2)E_{\bar D^*_l}(-p_2')}}\{[\varepsilon^{\varrho'}(Q_2)\cdot\varepsilon^{\vartheta'}(-p_2')]\bar{h}_{1}^{(\text{v})}(w^{2})(Q_2-p_2')_{\beta}\\
&-\bar{h}^{(\text{v})}_{2}(w^{2})\{[\varepsilon^{\varrho'}(Q_2)\cdot (-p_2')]\varepsilon^{\vartheta'}_{\beta}(-p_2')+[\varepsilon^{\vartheta'}(-p_2')\cdot Q_2]\varepsilon^{\varrho'}_{\beta}(Q_2)\}\},
\end{split}
\end{equation}
\end{subequations}
where $w=p-(Q_1-Q_2)/2$ is the momentum of the light meson, $h(w^2)$ and $\bar h(w^2)$ are the heavy meson form factors with respect to $\epsilon'$. Similarly, taking away the external lines including normalizations and polarization vectors $\varepsilon^\varrho_\nu(Q_1)$, $\varepsilon^\vartheta_\lambda(p_1')$, $\varepsilon^{\varrho'}_\mu(Q_2)$, $\varepsilon^{\vartheta'}_\tau(-p_2')$, we obtain the interaction from one light meson ($\sigma$, $\rho^0$, $V_1$ and $V_8$) exchange
\begin{equation}\label{kerneleps}
\begin{split}
\mathcal{V}&_{\nu\lambda,\tau\mu}(Q_1,Q_2,\mathbf{p})\\
=&-2E_1(\epsilon')F_1(\mathbf{w}^2)\frac{-ig_\sigma^2}{w^2+M_\sigma^2}2E_2(\epsilon')F_1(\mathbf{w}^2)\delta_{\nu\lambda}\delta_{\tau\mu}+F_2(\mathbf{w}^2)\left(\frac{-ig_\rho^2}{w^2+M_\rho^2}+\frac{-ig_1^2}{w^2+M_\omega^2}+\frac{-ig_8^2}{w^2+M_\phi^2}\right)\\
&\times F_2(\mathbf{w}^2)\{(Q_1+p_1')\cdot(Q_2-p_2')\delta_{\nu\lambda}\delta_{\tau\mu}-\delta_{\nu\lambda}[-(Q_1+p_1')_\tau p_{2\mu}'+Q_{2\tau}(Q_1+p_1')_\mu]\\
&-[p_{1\nu}'(Q_2-p_2')_\lambda+(Q_2-p_2')_\nu Q_{1\lambda}]\delta_{\tau\mu}-p_{1\nu}'\delta_{\lambda\tau}p_{2\mu}'+p_{1\nu}'\delta_{\lambda\mu}Q_{2\tau}-\delta_{\nu\tau}Q_{1\lambda}p_{2\mu}'+\delta_{\nu\mu}Q_{1\lambda}Q_{2\tau}\},
\end{split}
\end{equation}
where $E_1(\epsilon')=E_2(\epsilon')=\sqrt{\epsilon'M_0}/2$, and $w=(\mathbf{w},0)$. The interaction from one-$\rho^{\pm}$ exchange becomes
\begin{equation}\label{kerneleps'}
\begin{split}
\mathcal{V}&'_{\nu\lambda,\tau\mu}(Q_1,Q_2,\mathbf{p})\\
=&F_2(\mathbf{w}^2)\frac{-i2g_\rho^2}{w^2+M_\rho^2}F_2(\mathbf{w}^2)\{(Q_1+p_1')\cdot(Q_2-p_2')\delta_{\nu\lambda}\delta_{\tau\mu}-\delta_{\nu\lambda}[-(Q_1+p_1')_\tau p_{2\mu}'+Q_{2\tau}(Q_1+p_1')_\mu]\\
&-[p_{1\nu}'(Q_2-p_2')_\lambda+(Q_2-p_2')_\nu Q_{1\lambda}]\delta_{\tau\mu}-p_{1\nu}'\delta_{\lambda\tau}p_{2\mu}'+p_{1\nu}'\delta_{\lambda\mu}Q_{2\tau}-\delta_{\nu\tau}Q_{1\lambda}p_{2\mu}'+\delta_{\nu\mu}Q_{1\lambda}Q_{2\tau}\}.
\end{split}
\end{equation}
These momenta in Fig. \ref{Fig13} become
\begin{equation}\label{momenta3}
\begin{split}
w=(\mathbf{p}-\mathbf{Q}_{D^*}(\epsilon'),0),~~Q_1+Q_2=P^{\epsilon'},~~p_1'=p+P/2,~~p_2'=p-P/2,
\end{split}
\end{equation}
where $P=(0,0,0,iM_0)$, $P^{\epsilon'}=(0,0,0,i\epsilon')$, $Q_1=(\mathbf{Q}_{D^*}(\epsilon'),i\epsilon'/2)$, $Q_2=(-\mathbf{Q}_{D^*}(\epsilon'),i\epsilon'/2)$ and $\mathbf{Q}_{D^*}^2(\epsilon')=[\epsilon'^2-(M_{D^*_l}+M_{\bar D^*_l})^2]/4$.

Substituting Eqs. (\ref{kerneleps}) and (\ref{kerneleps'}) into (\ref{MEcurr3}), we obtain the explicit forms for tensors $\mathcal{M}^{c'_4}_{\nu\mu}$ and $\mathcal{M}'^{c'_4}_{\nu\mu}$. The $p$ integral is also computed in instantaneous approximation. $\mathcal{M}^{c'_4}_{\nu\mu}$ and $\mathcal{M}'^{c'_4}_{\nu\mu}$ only depend on $Q_1$ and $Q_2$, which can be calculated by means of the method given in Sec. \ref{sec:Jpsiobm}. Applying Eq. (\ref{Iepsilon'}), we obtain the function $\mathbb{I}_4(\epsilon')$ for closed channel $D^*\bar D^*$
\begin{equation}\label{Iepsilon'4}
\begin{split}
\mathbb{I}_4(\epsilon')&=\frac{1}{2}\int d^3Q_1d^3Q_2(2\pi)^4\delta^{(4)}(Q_1+Q_2-P^{\epsilon'})\sum_{\varrho'=1}^3\sum_{\varrho=1}^3|T_{(c'_4;b)a}(\epsilon')|^2.
\end{split}
\end{equation}

\section{physical mass and width of resonance}\label{sec:mw}
For resonance $\chi_{c0}(3915)$, the dispersion relation (\ref{disrel}) becomes
\begin{equation}
\begin{split}
\mathbb{D}(M_0)=&-\frac{\mathcal{P}}{\pi}\int_{\epsilon_{c'_1,M}}^\infty \frac{\mathbb{I}_1(\epsilon')}{\epsilon'-M_0}d\epsilon'-\frac{\mathcal{P}}{\pi}\int_{\epsilon_{c'_2,M}}^\infty \frac{\mathbb{I}_2(\epsilon')}{\epsilon'-M_0}d\epsilon'-\frac{\mathcal{P}}{\pi}\int_{\epsilon_{c'_3,M}}^\infty \frac{\mathbb{I}_3(\epsilon')}{\epsilon'-M_0}d\epsilon'\\
&-\frac{1}{\pi}\int_{\epsilon_{c'_4,M}}^\infty \frac{\mathbb{I}_4(\epsilon')}{\epsilon'-M_0}d\epsilon',
\end{split}
\end{equation}
where $\epsilon_{c'_1,M}=M_{J/\psi}+M_\omega$, $\epsilon_{c'_2,M}=M_{D^+}+M_{D^-}$, $\epsilon_{c'_3,M}=M_{D^0}+M_{\bar D^0}$, and $\epsilon_{c'_4,M}=M_{D_l^*}+M_{\bar D_l^*}$. From Eq. (\ref{pole}), we obtain that the physical mass of resonance $\chi_{c0}(3915)$ is $M=M_0+(2\pi)^3\mathbb{D}(M_0)$. Replacing $M_0$ by $M$ in Eqs. (\ref{Tmatrele}) and (\ref{decaywidthM0}), we recalculate the matrix element $T_{(c'_1;b)a}(M)$ and obtain the width $\Gamma_1$ for physical decay model $\chi_{c0}(3915)\rightarrow J/\psi\omega$. Replacing $M_0$ by $M$ in Eqs. (\ref{Tmatrele2}) and (\ref{decaywidth2M0}), we recalculate the matrix element $T_{(c'_2;b)a}(M)$ and obtain the width $\Gamma_2$ for physical decay model $\chi_{c0}(3915)\rightarrow D^+D^-$. For the isospin conservation, it is easy to obtain the width $\Gamma_3$ for physical decay model $\chi_{c0}(3915)\rightarrow D^0\bar D^0$.

\section{numerical result}\label{sec:nr}
Considering the isospin conservation, we employ the constituent quark masses $m_u=m_d=0.33$ GeV, the heavy quark mass $m_{c}=1.55$ GeV \cite{ts:Ebert} and the meson masses $M_\sigma=0.45$ GeV, $M_{\omega}=0.782$ GeV, $M_{\rho^0}=M_{\rho^{\pm}}=0.775$ GeV, $M_\phi=1.019$ GeV, $M_{D^{*0}}=M_{D^{*+}}=2.007$ GeV, $M_{D^{0}}=M_{D^{+}}=1.865$ GeV, $M_{J/\psi}=3.097$ GeV \cite{PDG2022}. Without an adjustable parameter, we numerically solve the eigenvalue equation (\ref{eigeneq}) and obtain the masses and wave functions of pure bound states $D^{*0}\bar{D}^{*0}$ and $D^{*+}D^{*-}$ with spin-parity quantum numbers $0^+$. Considering the cross terms between these two pure bound states $D^{*0}\bar{D}^{*0}$ and $D^{*+}D^{*-}$ and using the coupled-channel approach, we obtain the mass $M_0$ of the mixed state with $0^+$. Then $M_0$ and GBS wave function $\chi^{D^{*}\bar D^{*},0^+}(P,p,k,k')$ given in Eq. (\ref{mmbounstate1}) are used to evaluate the matrix elements $T_{(c'_1;b)a}(M_0)$ and $T_{(c'_2;b)a}(M_0)$ with mass of meson-meson bound state, and the decay widths $\Gamma_1(M_0)$ and $\Gamma_2(M_0)$ with mass of meson-meson bound state should not be the width of physical resonance. From Eqs. (\ref{Tmatrele}), (\ref{MEcurr}) and (\ref{momenta}), we calculate the $T$-matrix element $T_{(c'_1;b)a}(\epsilon')$ with respect to $\epsilon'$ for channel $J/\psi\omega$. From Eqs. (\ref{kernel2}), (\ref{kernel2'}), (\ref{Tmatrele2}), (\ref{MEcurr2}) and (\ref{momenta2}), we calculate the $T$-matrix element $T_{(c'_2;b)a}(\epsilon')$ with respect to $\epsilon'$ for channel $D^+D^-$. From Eqs. (\ref{Tmatrele3}), (\ref{MEcurr3}), (\ref{kerneleps}), (\ref{kerneleps'}) and (\ref{momenta3}), we calculate the $T$-matrix element $T_{(c'_4;b)a}(\epsilon')$ with respect to $\epsilon'$ for closed channel $D^*\bar D^*$. From Eqs. (\ref{Iepsilon'1}), (\ref{Iepsilon'2}) and (\ref{Iepsilon'4}), we calculate the functions $\mathbb{I}_1(\epsilon')$ over $\epsilon_{c'_1,M}<\epsilon'<\infty$, $\mathbb{I}_2(\epsilon')$ over $\epsilon_{c'_2,M}<\epsilon'<\infty$, $\mathbb{I}_3(\epsilon')$ over $\epsilon_{c'_3,M}<\epsilon'<\infty$, and $\mathbb{I}_4(\epsilon')$ over $\epsilon_{c'_4,M}<\epsilon'<\infty$, respectively. By doing the numerical calculation, we obtain the mass correction $\Delta M=(2\pi)^3\mathbb{D}(M_0)$ and the physical mass $M$ for resonance $\chi_{c0}(3915)$. Finally, the physical mass is used to recalculate these strong decay widths $\Gamma_1(\chi_{c0}(3915)\rightarrow J/\psi\omega)$, $\Gamma_2(\chi_{c0}(3915)\rightarrow D^+D^-)$~ and $\Gamma_3(\chi_{c0}(3915)\rightarrow D^0\bar D^0)$. Some errors in Ref. \cite{mypaper8} have been revised. $M$ and $\Gamma$ should be the observed mass and full width in experiments. Our numerical results for resonance $\chi_{c0}(3915)$ are in good agreement with the experimental data, which are presented in Table \ref{table1}.
\newcommand{\tabincell}[2]{\begin{tabular}{@{}#1@{}}#2\end{tabular}}
\begin{table}[ht]
\caption{Mass $M$ and full width $\Gamma$ for physical resonance $\chi_{c0}(3915)$. $M_0$ is the mass of mixed state of two bound states $D^{*0}\bar{D}^{*0}$ and $D^{*+}D^{*-}$, $\Delta M$ is the calculated correction due to all open and closed channels, and $\Gamma_i$ is the calculated width of $i$th decay channel. (Dimensioned quantities in MeV.)}
\label{table1} \vspace*{-6pt}
\begin{center}
\begin{tabular}{cccccccc}\hline
Quantity  & ~~$M_0$ & ~~$\Delta M$ & ~~$M$ & ~~~$\Gamma_1$ & ~~~$\Gamma_2$ & ~~~$\Gamma_3$ & ~~$\Gamma$
\\ \hline
this work     & ~~3953.7 & ~~$-$30.9 &  ~~3922.8  & ~~~22.5 & ~~~1.5 &~~~1.5&~~25.5
\\
\hline
PDG\cite{PDG2022} & & & ~~3921.7$\pm$1.8 & & & & ~~18.8$\pm$3.5
\\
\hline
\end{tabular}
\end{center}
\end{table}

It is necessary to emphasize that there is not an adjustable parameter in our approach. We require the meson-quark coupling constants $g$ and the parameters $\omega_H$ in BS amplitudes of heavy mesons to calculate the mass and decay width of physical resonance. The meson-quark coupling constants can be determined by QCD sum rules approach \cite{cc2}, and these parameters in BS amplitudes of heavy mesons are fixed by providing fits to observables \cite{BSE:Roberts4,BSE:Roberts5,Jpsi1}. Our approach also involves the constituent quark masses $m_u$, $m_d$, and the heavy quark mass $m_c$. According to the spontaneous breaking of chiral symmetry, the light quarks ($u,d,s$) obtain their constituent masses because the vacuum condensate is not equal to zero, and the heavy quark mass $m_c$ is irrelevant to vacuum condensate. Normally, the value slightly greater than a third of nucleon mass is employed as the constituent mass of light quark. The value of heavy quark mass $m_c$ can be determined by the experimental mass of charmonium system $J/\psi$. Of course, the values of these parameters, including $g$, $\omega_H$, $m_u$, $m_d$ and $m_c$, are values in respective ranges. Simultaneously varying these parameters in respective ranges, we find that the uncertainties of numerical results are at most 5\%. Despite the large uncertainty of meson mass $M_\sigma$, it has been found that the uncertainties of numerical results from meson mass $M_\sigma$ are also very small in our previous works \cite{mypaper4,mypaper5,mypaper6,mypaper7} and Refs. \cite{Msigma1,Msigma2}. Therefore, in our approach the calculated mass and decay width are uniquely determined.

Up to now, a theoretical approach from QCD to investigate resonance which is regarded as an unstable two-body system has been established. In this paper, we only explore exotic meson resonance which is considered as an unstable molecular state composed of two heavy vector mesons. The extension of our approach to more general resonances is straightforward, while the interaction Lagrangian may be modified. More importantly, it is most reasonable and fascinating to investigate resonance as far as possible from QCD. In the framework of quantum field theory, the nonperturbative contribution from the vacuum condensates can be introduced into BS wave function \cite{mypaper5} and the two-particle irreducible Green's function, and then the calculated mass and decay width of resonance will contain more inspiration of QCD.

\section{conclusion}\label{sec:concl}
Exotic resonance $\chi_{c0}(3915)$ is considered as a mixed state of two unstable molecular states $D^{*0}\bar{D}^{*0}$ and $D^{*+}D^{*-}$, and we investigate the time evolution of meson-meson molecular state as determined by the total Hamiltonian. According to the developed Bethe-Salpeter theory, the total matrix elements for all decay channels should be calculated with respect to arbitrary value of the final state energy. Because the total energy of the final state extends from $\epsilon_M$ to $+\infty$, we consider three open decay channels $J/\psi\omega$, $D^+D^-$, $D^0\bar D^0$ and one closed channel $D^{*}\bar D^{*}$ from the effective interaction Lagrangian at low energy QCD, which are exhibited by extended Feynman diagrams. Using the developed Bethe-Salpeter theory, we calculate the mass $M$ and full width $\Gamma$ of physical resonance $\chi_{c0}(3915)$, which are in good agreement with the experimental data. Obviously, our work can be extended to more general resonances.

\begin{acknowledgements}
This work was supported by the National Natural Science Foundation of China under Grants No. 11705104, 11801323 and 52174145; Shandong Provincial Natural Science Foundation, China under Grants No. ZR2016AQ19 and ZR2016AM31; and SDUST Research Fund under Grant No. 2018TDJH101.
\end{acknowledgements}

\appendix

\section{tensor structures in the general form of BS wave functions}\label{app1}
The tensor structures in Eqs. (\ref{jp}) and (\ref{jm}) are given below \cite{mypaper4,mypaper7}
\begin{equation*}
\mathcal{T}_{\lambda\tau}^1=(p^2+\eta_1P\cdot p-\eta_2P\cdot p-\eta_1\eta_2P^2)g_{\lambda\tau}-(p_{\lambda}p_{\tau}+\eta_1P_{\tau}p_{\lambda}-\eta_2P_{\lambda}p_{\tau}-\eta_1\eta_2P_{\lambda}P_{\tau}),
\end{equation*}
\begin{equation*}
\begin{split}
\mathcal{T}_{\lambda\tau}^2=&(p^2+2\eta_1P\cdot p+\eta_1^2P^2)(p^2-2\eta_2P\cdot p+\eta_2^2P^2)g_{\lambda\tau}\\
&+(p^2+\eta_1P\cdot p-\eta_2P\cdot p-\eta_1\eta_2P^2)(p_{\lambda}p_{\tau}+\eta_1P_{\lambda}p_{\tau}-\eta_2P_{\tau}p_{\lambda}-\eta_1\eta_2P_{\lambda}P_{\tau})\\
&-(p^2-2\eta_2P\cdot p+\eta_2^2P^2)(p_{\lambda}p_{\tau}+\eta_1P_{\lambda}p_{\tau}+\eta_1P_{\tau}p_{\lambda}+\eta_1 ^2P_{\lambda}P_{\tau})\\
&-(p^2+2\eta_1P\cdot p+\eta_1^2P^2)(p_{\lambda}p_{\tau}-\eta_2P_{\lambda}p_{\tau}-\eta_2P_{\tau}p_{\lambda}+\eta_2 ^2P_{\lambda}P_{\tau}),
\end{split}
\end{equation*}
\begin{equation*}
\begin{split}
\mathcal{T}_{\mu_1\cdots\mu_j\lambda\tau}^3=&\frac{1}{j!}p_{\{\mu_2}\cdots p_{\mu_j}g_{\mu_1\}\lambda}(p^2+2\eta_1P\cdot p+\eta_1^2P^2)[(p^2-2\eta_2P\cdot p+\eta_2^2P^2)(p+\eta_1P)_{\tau}\\
&-(p^2+\eta_1P\cdot p-\eta_2P\cdot p-\eta_1\eta_2P^2)(p-\eta_2P)_{\tau}]\\
&-p_{\mu_1}\cdots p_{\mu_j}[(p^2-2\eta_2P\cdot p+\eta_2^2P^2)(p_{\lambda}p_{\tau}+\eta_1P_{\lambda}p_{\tau}+\eta_1P_{\tau}p_{\lambda}+\eta_1 ^2P_{\lambda}P_{\tau})\\
&-(p^2+\eta_1P\cdot p-\eta_2P\cdot p-\eta_1\eta_2P^2)(p_{\lambda}p_{\tau}+\eta_1P_{\lambda}p_{\tau}-\eta_2P_{\tau}p_{\lambda}-\eta_1\eta_2P_{\lambda}P_{\tau})],
\end{split}
\end{equation*}
\begin{equation*}
\begin{split}
\mathcal{T}_{\mu_1\cdots\mu_j\lambda\tau}^4=&\frac{1}{j!}p_{\{\mu_2}\cdots p_{\mu_j}g_{\mu_1\}\tau}(p^2-2\eta_2P\cdot p+\eta_2^2P^2)[(p^2+\eta_1P\cdot p\\
&-\eta_2P\cdot p-\eta_1\eta_2P^2)(p+\eta_1P)_{\lambda}-(p^2+2\eta_1P\cdot p+\eta_1^2P^2)(p-\eta_2P)_{\lambda}]\\
&+p_{\mu_1}\cdots p_{\mu_j}[(p^2+2\eta_1P\cdot p+\eta_1^2P^2)(p_{\lambda}p_{\tau}-\eta_2P_{\lambda}p_{\tau}-\eta_2P_{\tau}p_{\lambda}+\eta_2 ^2P_{\lambda}P_{\tau})\\
&-(p^2+\eta_1P\cdot p-\eta_2P\cdot p-\eta_1\eta_2P^2)(p_{\lambda}p_{\tau}+\eta_1P_{\lambda}p_{\tau}-\eta_2P_{\tau}p_{\lambda}-\eta_1\eta_2P_{\lambda}P_{\tau})],
\end{split}
\end{equation*}
\begin{equation*}
\begin{split}
\mathcal{T}^5_{\mu_1\cdots\mu_j\lambda\tau}=&\frac{1}{j!}(p^2+2\eta_1P\cdot p+\eta_1^2P^2)(p^2-2\eta_2P\cdot p+\eta_2^2P^2)p_{\{\mu_3}\cdots p_{\mu_j}g_{\mu_1\lambda}g_{\mu_2\}\tau}\\
&-\frac{1}{j!}p_{\{\mu_2}\cdots p_{\mu_j}g_{\mu_1\}\tau}(p^2-2\eta_2P\cdot p+\eta_2^2P^2)(p+\eta_1P)_{\lambda}\\
&-\frac{1}{j!}p_{\{\mu_2}\cdots p_{\mu_j}g_{\mu_1\}\lambda}(p^2+2\eta_1P\cdot p+\eta_1^2P^2)(p-\eta_2P)_{\tau}\\
&+p_{\mu_1}\cdots p_{\mu_j}(p_{\lambda}p_{\tau}+\eta_1P_{\lambda}p_{\tau}-\eta_2P_{\tau}p_{\lambda}-\eta_1\eta_2P_{\lambda}P_{\tau}),
\end{split}
\end{equation*}
\begin{equation*}
\begin{split}
\mathcal{T}^6_{\mu_1\cdots\mu_j\lambda\tau}=&p_{\{\mu_3}\cdots p_{\mu_j}\epsilon_{\mu_{1}\lambda\xi\zeta}p_\xi P_\zeta\epsilon_{\mu_{2}\}\tau\xi'\zeta'}p_{\xi'}P_{\zeta'},
\end{split}
\end{equation*}
\begin{equation*}
\begin{split}
\mathcal{T}_{\mu_1\cdots\mu_j\lambda\tau}^7=&-(2p^2+\eta_1P\cdot p-\eta_2P\cdot p)p_{\{\mu_2}\cdots p_{\mu_j}\epsilon_{\mu_1\}\lambda\tau\xi}p_\xi\\
&+(2\eta_1\eta_2P\cdot p+\eta_2p^2-\eta_1p^2)p_{\{\mu_2}\cdots p_{\mu_j}\epsilon_{\mu_1\}\lambda\tau\xi}P_\xi\\
&+p_{\{\mu_2}\cdots p_{\mu_j}\epsilon_{\mu_1\}\lambda\xi\zeta}p_\xi P_\zeta p_\tau+p_{\{\mu_2}\cdots p_{\mu_j}\epsilon_{\mu_1\}\tau\xi\zeta}p_\xi P_\zeta p_\lambda,
\end{split}
\end{equation*}
\begin{equation*}\
\begin{split}
\mathcal{T}_{\mu_1\cdots\mu_j\lambda\tau}^8=&-(P\cdot p)p_{\{\mu_2}\cdots p_{\mu_j}\epsilon_{\mu_1\}\lambda\tau\xi}p_\xi+p^2p_{\{\mu_2}\cdots p_{\mu_j}\epsilon_{\mu_1\}\lambda\tau\xi}P_\xi\\
&-p_{\{\mu_2}\cdots p_{\mu_j}\epsilon_{\mu_1\}\lambda\xi\zeta}p_\xi P_\zeta p_\tau+p_{\{\mu_2}\cdots p_{\mu_j}\epsilon_{\mu_1\}\tau\xi\zeta}p_\xi P_\zeta p_\lambda,
\end{split}
\end{equation*}
\begin{equation*}
\begin{split}
\mathcal{T}_{\mu_1\cdots\mu_j\lambda\tau}^{9}=&-(2P\cdot p+\eta_1P^2-\eta_2P^2)p_{\{\mu_2}\cdots p_{\mu_j}\epsilon_{\mu_1\}\lambda\tau\xi}p_\xi\\
&+P\cdot(\eta_2p-\eta_1P+2\eta_1\eta_2P)p_{\{\mu_2}\cdots p_{\mu_j}\epsilon_{\mu_1\}\lambda\tau\xi}P_\xi\\
&+p_{\{\mu_2}\cdots p_{\mu_j}\epsilon_{\mu_1\}\lambda\xi\zeta}p_\xi P_\zeta P_\tau+p_{\{\mu_2}\cdots p_{\mu_j}\epsilon_{\mu_1\}\tau\xi\zeta}p_\xi P_\zeta P_\lambda,
\end{split}
\end{equation*}
\begin{equation*}
\begin{split}
\mathcal{T}_{\mu_1\cdots\mu_j\lambda\tau}^{10}=&-P^2p_{\{\mu_2}\cdots p_{\mu_j}\epsilon_{\mu_1\}\lambda\tau\xi}p_\xi+(P\cdot p)p_{\{\mu_2}\cdots p_{\mu_j}\epsilon_{\mu_1\}\lambda\tau\xi}P_\xi\\
&-p_{\{\mu_2}\cdots p_{\mu_j}\epsilon_{\mu_1\}\lambda\xi\zeta}p_\xi P_\zeta P_\tau+p_{\{\mu_2}\cdots p_{\mu_j}\epsilon_{\mu_1\}\tau\xi\zeta}p_\xi P_\zeta P_\lambda,
\end{split}
\end{equation*}
\begin{equation*}
\begin{split}
\mathcal{T}_{\mu_1\cdots\mu_j\lambda\tau}^{11}=&(p^2+\eta_1P\cdot p-\eta_2P\cdot p-\eta_1\eta_2P^2)p_{\{\mu_3}\cdots p_{\mu_j}g_{\mu_1\lambda}\epsilon_{\mu_2\}\tau\xi\zeta}p_\xi P_\zeta\\
&-p_{\{\mu_2}\cdots p_{\mu_j}\epsilon_{\mu_1\}\tau\xi\zeta}p_\xi P_\zeta(p-\eta_2P)_\lambda,
\end{split}
\end{equation*}
\begin{equation*}
\begin{split}
\mathcal{T}_{\mu_1\cdots\mu_j\lambda\tau}^{12}=&(p^2+\eta_1P\cdot p-\eta_2P\cdot p-\eta_1\eta_2P^2)p_{\{\mu_3}\cdots p_{\mu_j}g_{\mu_1\tau}\epsilon_{\mu_2\}\lambda\xi\zeta}p_\xi P_\zeta\\
&-p_{\{\mu_2}\cdots p_{\mu_j}\epsilon_{\mu_1\}_\lambda\xi\zeta}p_\xi P_\zeta(p+\eta_1P)_\tau.
\end{split}
\end{equation*}

\bibliographystyle{apsrev}
\bibliography{ref}

\begin{thebibliography}{33}
\expandafter\ifx\csname natexlab\endcsname\relax\def\natexlab#1{#1}\fi
\expandafter\ifx\csname bibnamefont\endcsname\relax
  \def\bibnamefont#1{#1}\fi
\expandafter\ifx\csname bibfnamefont\endcsname\relax
  \def\bibfnamefont#1{#1}\fi
\expandafter\ifx\csname citenamefont\endcsname\relax
  \def\citenamefont#1{#1}\fi
\expandafter\ifx\csname url\endcsname\relax
  \def\url#1{\texttt{#1}}\fi
\expandafter\ifx\csname urlprefix\endcsname\relax\def\urlprefix{URL }\fi
\providecommand{\bibinfo}[2]{#2}
\providecommand{\eprint}[2][]{\url{#2}}

\bibitem[{\citenamefont{Swanson}(2004)}]{ms:Swanson}
\bibinfo{author}{\bibfnamefont{E.~S.} \bibnamefont{Swanson}},
  \bibinfo{journal}{Phys. Lett. B} \textbf{\bibinfo{volume}{588}},
  \bibinfo{pages}{189} (\bibinfo{year}{2004}).

\bibitem[{\citenamefont{T{\"o}rnqvist}(2004)}]{ms:Torn}
\bibinfo{author}{\bibfnamefont{N.~A.} \bibnamefont{T{\"o}rnqvist}},
  \bibinfo{journal}{Phys. Lett. B} \textbf{\bibinfo{volume}{590}},
  \bibinfo{pages}{209} (\bibinfo{year}{2004}).

\bibitem[{\citenamefont{Liu and Zhu}(2009)}]{liu}
\bibinfo{author}{\bibfnamefont{X.}~\bibnamefont{Liu}} \bibnamefont{and}
  \bibinfo{author}{\bibfnamefont{S.~L.} \bibnamefont{Zhu}},
  \bibinfo{journal}{Phys. Rev. D} \textbf{\bibinfo{volume}{80}},
  \bibinfo{pages}{017502} (\bibinfo{year}{2009}).

\bibitem[{\citenamefont{Chen and L{\"u}}(2015)}]{mypaper4}
\bibinfo{author}{\bibfnamefont{X.}~\bibnamefont{Chen}} \bibnamefont{and}
  \bibinfo{author}{\bibfnamefont{X.}~\bibnamefont{L{\"u}}},
  \bibinfo{journal}{Eur. Phys. J. C} \textbf{\bibinfo{volume}{75}},
  \bibinfo{pages}{98} (\bibinfo{year}{2015}).

\bibitem[{\citenamefont{Zhao et~al.}(2022)\citenamefont{Zhao, Wang, Wang, and
  Guo}}]{Msigma2}
\bibinfo{author}{\bibfnamefont{M.-J.} \bibnamefont{Zhao}},
  \bibinfo{author}{\bibfnamefont{Z.-Y.} \bibnamefont{Wang}},
  \bibinfo{author}{\bibfnamefont{C.}~\bibnamefont{Wang}}, \bibnamefont{and}
  \bibinfo{author}{\bibfnamefont{X.-H.} \bibnamefont{Guo}},
  \bibinfo{journal}{Phys. Rev. D} \textbf{\bibinfo{volume}{105}},
  \bibinfo{pages}{096016} (\bibinfo{year}{2022}).

\bibitem[{\citenamefont{Chen and L\"u}(2018)}]{mypaper6}
\bibinfo{author}{\bibfnamefont{X.}~\bibnamefont{Chen}} \bibnamefont{and}
  \bibinfo{author}{\bibfnamefont{X.}~\bibnamefont{L\"u}},
  \bibinfo{journal}{Phys. Rev. D} \textbf{\bibinfo{volume}{97}},
  \bibinfo{pages}{114005} (\bibinfo{year}{2018}).

\bibitem[{\citenamefont{Chen et~al.}(2020)\citenamefont{Chen, L\"u, Shi, Guo,
  and Wang}}]{mypaper7}
\bibinfo{author}{\bibfnamefont{X.}~\bibnamefont{Chen}},
  \bibinfo{author}{\bibfnamefont{X.}~\bibnamefont{L\"u}},
  \bibinfo{author}{\bibfnamefont{R.}~\bibnamefont{Shi}},
  \bibinfo{author}{\bibfnamefont{X.}~\bibnamefont{Guo}}, \bibnamefont{and}
  \bibinfo{author}{\bibfnamefont{Q.}~\bibnamefont{Wang}},
  \bibinfo{journal}{Phys. Rev. D} \textbf{\bibinfo{volume}{101}},
  \bibinfo{pages}{014009} (\bibinfo{year}{2020}).

\bibitem[{\citenamefont{Chen and L{\"u}}(2023)}]{mypaper8}
\bibinfo{author}{\bibfnamefont{X.}~\bibnamefont{Chen}} \bibnamefont{and}
  \bibinfo{author}{\bibfnamefont{X.}~\bibnamefont{L{\"u}}},
  \bibinfo{journal}{Eur. Phys. J. C} \textbf{\bibinfo{volume}{83}},
  \bibinfo{pages}{499} (\bibinfo{year}{2023}).

\bibitem[{\citenamefont{Esposito et~al.}(2017)\citenamefont{Esposito, Pilloni,
  and Polosa}}]{mutlistate}
\bibinfo{author}{\bibfnamefont{A.}~\bibnamefont{Esposito}},
  \bibinfo{author}{\bibfnamefont{A.}~\bibnamefont{Pilloni}}, \bibnamefont{and}
  \bibinfo{author}{\bibfnamefont{A.~D.} \bibnamefont{Polosa}},
  \bibinfo{journal}{Phys. Rep.} \textbf{\bibinfo{volume}{668}},
  \bibinfo{pages}{1} (\bibinfo{year}{2017}).

\bibitem[{\citenamefont{Chen et~al.}(2013)\citenamefont{Chen, Liu, Shi, and
  L\"u}}]{mypaper3}
\bibinfo{author}{\bibfnamefont{X.}~\bibnamefont{Chen}},
  \bibinfo{author}{\bibfnamefont{R.}~\bibnamefont{Liu}},
  \bibinfo{author}{\bibfnamefont{R.}~\bibnamefont{Shi}}, \bibnamefont{and}
  \bibinfo{author}{\bibfnamefont{X.}~\bibnamefont{L\"u}},
  \bibinfo{journal}{Phys. Rev. D} \textbf{\bibinfo{volume}{87}},
  \bibinfo{pages}{065013} (\bibinfo{year}{2013}).

\bibitem[{\citenamefont{Chen et~al.}(2016)\citenamefont{Chen, L{\"u}, Shi, and
  Guo}}]{mypaper5}
\bibinfo{author}{\bibfnamefont{X.}~\bibnamefont{Chen}},
  \bibinfo{author}{\bibfnamefont{X.}~\bibnamefont{L{\"u}}},
  \bibinfo{author}{\bibfnamefont{R.}~\bibnamefont{Shi}}, \bibnamefont{and}
  \bibinfo{author}{\bibfnamefont{X.}~\bibnamefont{Guo}},
  \bibinfo{journal}{Nucl. Phys. B} \textbf{\bibinfo{volume}{909}},
  \bibinfo{pages}{243 } (\bibinfo{year}{2016}).

\bibitem[{\citenamefont{Ambrosino et~al.}(2009)}]{mix}
\bibinfo{author}{\bibfnamefont{F.}~\bibnamefont{Ambrosino}}
  \bibnamefont{et~al.} (\bibinfo{collaboration}{KLOE Collaboration}),
  \bibinfo{journal}{JHEP} \textbf{\bibinfo{volume}{07}}, \bibinfo{pages}{105}
  (\bibinfo{year}{2009}).

\bibitem[{\citenamefont{Reinders et~al.}(1985)\citenamefont{Reinders,
  Rubinstein, and Yazaki}}]{cc2}
\bibinfo{author}{\bibfnamefont{L.}~\bibnamefont{Reinders}},
  \bibinfo{author}{\bibfnamefont{H.}~\bibnamefont{Rubinstein}},
  \bibnamefont{and} \bibinfo{author}{\bibfnamefont{S.}~\bibnamefont{Yazaki}},
  \bibinfo{journal}{Phys. Rep.} \textbf{\bibinfo{volume}{127}},
  \bibinfo{pages}{1} (\bibinfo{year}{1985}).

\bibitem[{\citenamefont{L{\"u} et~al.}(1996)\citenamefont{L{\"u}, Liu, and
  Zhao}}]{cc1a}
\bibinfo{author}{\bibfnamefont{X.}~\bibnamefont{L{\"u}}},
  \bibinfo{author}{\bibfnamefont{Y.}~\bibnamefont{Liu}}, \bibnamefont{and}
  \bibinfo{author}{\bibfnamefont{E.}~\bibnamefont{Zhao}},
  \bibinfo{journal}{Chinese Physics Letters} \textbf{\bibinfo{volume}{13}},
  \bibinfo{pages}{652} (\bibinfo{year}{1996}).

\bibitem[{\citenamefont{L{\"u} et~al.}(1997)\citenamefont{L{\"u}, Liu, and
  Zhao}}]{cc1b}
\bibinfo{author}{\bibfnamefont{X.}~\bibnamefont{L{\"u}}},
  \bibinfo{author}{\bibfnamefont{Y.}~\bibnamefont{Liu}}, \bibnamefont{and}
  \bibinfo{author}{\bibfnamefont{E.}~\bibnamefont{Zhao}},
  \bibinfo{journal}{Science in China (Series A)} \textbf{\bibinfo{volume}{27}},
  \bibinfo{pages}{361} (\bibinfo{year}{1997}).

\bibitem[{\citenamefont{Burden et~al.}(1997)\citenamefont{Burden, Qian,
  Roberts, Tandy, and Thomson}}]{BSE:Roberts1}
\bibinfo{author}{\bibfnamefont{C.~J.} \bibnamefont{Burden}},
  \bibinfo{author}{\bibfnamefont{L.}~\bibnamefont{Qian}},
  \bibinfo{author}{\bibfnamefont{C.~D.} \bibnamefont{Roberts}},
  \bibinfo{author}{\bibfnamefont{P.~C.} \bibnamefont{Tandy}}, \bibnamefont{and}
  \bibinfo{author}{\bibfnamefont{M.~J.} \bibnamefont{Thomson}},
  \bibinfo{journal}{Phys. Rev. C} \textbf{\bibinfo{volume}{55}},
  \bibinfo{pages}{2649} (\bibinfo{year}{1997}).

\bibitem[{\citenamefont{Maris et~al.}(1998)\citenamefont{Maris, Roberts, and
  Tandy}}]{BSE:Roberts3}
\bibinfo{author}{\bibfnamefont{P.}~\bibnamefont{Maris}},
  \bibinfo{author}{\bibfnamefont{C.~D.} \bibnamefont{Roberts}},
  \bibnamefont{and} \bibinfo{author}{\bibfnamefont{P.~C.} \bibnamefont{Tandy}},
  \bibinfo{journal}{Phys. Lett. B} \textbf{\bibinfo{volume}{420}},
  \bibinfo{pages}{267} (\bibinfo{year}{1998}).

\bibitem[{\citenamefont{Ivanov et~al.}(1999)\citenamefont{Ivanov, Kalinovsky,
  and Roberts}}]{BSE:Roberts4}
\bibinfo{author}{\bibfnamefont{M.~A.} \bibnamefont{Ivanov}},
  \bibinfo{author}{\bibfnamefont{Y.~L.} \bibnamefont{Kalinovsky}},
  \bibnamefont{and} \bibinfo{author}{\bibfnamefont{C.~D.}
  \bibnamefont{Roberts}}, \bibinfo{journal}{Phys. Rev. D}
  \textbf{\bibinfo{volume}{60}}, \bibinfo{pages}{034018}
  (\bibinfo{year}{1999}).

\bibitem[{\citenamefont{Ivanov et~al.}(2007)\citenamefont{Ivanov, K\"orner,
  Kovalenko, and Roberts}}]{BSE:Roberts5}
\bibinfo{author}{\bibfnamefont{M.~A.} \bibnamefont{Ivanov}},
  \bibinfo{author}{\bibfnamefont{J.~G.} \bibnamefont{K\"orner}},
  \bibinfo{author}{\bibfnamefont{S.~G.} \bibnamefont{Kovalenko}},
  \bibnamefont{and} \bibinfo{author}{\bibfnamefont{C.~D.}
  \bibnamefont{Roberts}}, \bibinfo{journal}{Phys. Rev. D}
  \textbf{\bibinfo{volume}{76}}, \bibinfo{pages}{034018}
  (\bibinfo{year}{2007}).

\bibitem[{\citenamefont{Neubert}(1994)}]{hqet}
\bibinfo{author}{\bibfnamefont{M.}~\bibnamefont{Neubert}},
  \bibinfo{journal}{Phys. Rep.} \textbf{\bibinfo{volume}{245}},
  \bibinfo{pages}{259} (\bibinfo{year}{1994}).

\bibitem[{\citenamefont{Chen et~al.}(2009)\citenamefont{Chen, Wang, Li, Zeng,
  Yu, and L{\"u}}}]{mypaper}
\bibinfo{author}{\bibfnamefont{X.}~\bibnamefont{Chen}},
  \bibinfo{author}{\bibfnamefont{B.}~\bibnamefont{Wang}},
  \bibinfo{author}{\bibfnamefont{X.}~\bibnamefont{Li}},
  \bibinfo{author}{\bibfnamefont{X.}~\bibnamefont{Zeng}},
  \bibinfo{author}{\bibfnamefont{S.}~\bibnamefont{Yu}}, \bibnamefont{and}
  \bibinfo{author}{\bibfnamefont{X.}~\bibnamefont{L{\"u}}},
  \bibinfo{journal}{Phys. Rev. D} \textbf{\bibinfo{volume}{79}},
  \bibinfo{pages}{114006} (\bibinfo{year}{2009}).

\bibitem[{\citenamefont{Choi et~al.}(2005)}]{Y39401}
\bibinfo{author}{\bibfnamefont{S.-K.} \bibnamefont{Choi}} \bibnamefont{et~al.}
  (\bibinfo{collaboration}{Belle Collaboration}), \bibinfo{journal}{Phys. Rev.
  Lett.} \textbf{\bibinfo{volume}{94}}, \bibinfo{pages}{182002}
  (\bibinfo{year}{2005}).

\bibitem[{\citenamefont{Aubert et~al.}(2008)}]{Y39402}
\bibinfo{author}{\bibfnamefont{B.}~\bibnamefont{Aubert}} \bibnamefont{et~al.}
  (\bibinfo{collaboration}{BABAR Collaboration}), \bibinfo{journal}{Phys. Rev.
  Lett.} \textbf{\bibinfo{volume}{101}}, \bibinfo{pages}{082001}
  (\bibinfo{year}{2008}).

\bibitem[{\citenamefont{Uehara et~al.}(2010)}]{X39151}
\bibinfo{author}{\bibfnamefont{S.}~\bibnamefont{Uehara}} \bibnamefont{et~al.}
  (\bibinfo{collaboration}{Belle Collaboration}), \bibinfo{journal}{Phys. Rev.
  Lett.} \textbf{\bibinfo{volume}{104}}, \bibinfo{pages}{092001}
  (\bibinfo{year}{2010}).

\bibitem[{\citenamefont{Lees et~al.}(2012)}]{Y39404}
\bibinfo{author}{\bibfnamefont{J.~P.} \bibnamefont{Lees}} \bibnamefont{et~al.}
  (\bibinfo{collaboration}{BABAR Collaboration}), \bibinfo{journal}{Phys. Rev.
  D} \textbf{\bibinfo{volume}{86}}, \bibinfo{pages}{072002}
  (\bibinfo{year}{2012}).

\bibitem[{\citenamefont{Vinokurova et~al.}(2015)}]{X39153}
\bibinfo{author}{\bibfnamefont{A.}~\bibnamefont{Vinokurova}}
  \bibnamefont{et~al.} (\bibinfo{collaboration}{Belle Collaboration}),
  \bibinfo{journal}{JHEP} \textbf{\bibinfo{volume}{06}}, \bibinfo{pages}{132}
  (\bibinfo{year}{2015}).

\bibitem[{\citenamefont{Aaij et~al.}(2020)}]{X39152}
\bibinfo{author}{\bibfnamefont{R.}~\bibnamefont{Aaij}} \bibnamefont{et~al.}
  (\bibinfo{collaboration}{LHCb Collaboration}), \bibinfo{journal}{Phys. Rev.
  D} \textbf{\bibinfo{volume}{102}}, \bibinfo{pages}{112003}
  (\bibinfo{year}{2020}).

\bibitem[{\citenamefont{Luri\'{e}}(1968)}]{ParticlesFields}
\bibinfo{author}{\bibfnamefont{D.}~\bibnamefont{Luri\'{e}}},
  \emph{\bibinfo{title}{{Particles and Fields}}}
  (\bibinfo{publisher}{Interscience Publishers, New York},
  \bibinfo{year}{1968}).

\bibitem[{\citenamefont{Goldberger and Watson}(1964)}]{GreenFun}
\bibinfo{author}{\bibfnamefont{M.~L.} \bibnamefont{Goldberger}}
  \bibnamefont{and} \bibinfo{author}{\bibfnamefont{K.~M.}
  \bibnamefont{Watson}}, \emph{\bibinfo{title}{{Collision Theory}}}
  (\bibinfo{publisher}{Wiley, New York}, \bibinfo{year}{1964}).

\bibitem[{\citenamefont{Ebert et~al.}(2006)\citenamefont{Ebert, Faustov, and
  Galkin}}]{ts:Ebert}
\bibinfo{author}{\bibfnamefont{D.}~\bibnamefont{Ebert}},
  \bibinfo{author}{\bibfnamefont{R.~N.} \bibnamefont{Faustov}},
  \bibnamefont{and} \bibinfo{author}{\bibfnamefont{V.~O.}
  \bibnamefont{Galkin}}, \bibinfo{journal}{Phys. Lett. B}
  \textbf{\bibinfo{volume}{634}}, \bibinfo{pages}{214} (\bibinfo{year}{2006}).

\bibitem[{\citenamefont{Workman et~al.}(2022)}]{PDG2022}
\bibinfo{author}{\bibfnamefont{R.~L.} \bibnamefont{Workman}}
  \bibnamefont{et~al.} (\bibinfo{collaboration}{Particle Data Group}),
  \bibinfo{journal}{Prog. Theor. Exp. Phys.} \textbf{\bibinfo{volume}{2022}},
  \bibinfo{pages}{083C01} (\bibinfo{year}{2022}).

\bibitem[{\citenamefont{Kawanai and Sasaki}(2011)}]{Jpsi1}
\bibinfo{author}{\bibfnamefont{T.}~\bibnamefont{Kawanai}} \bibnamefont{and}
  \bibinfo{author}{\bibfnamefont{S.}~\bibnamefont{Sasaki}},
  \bibinfo{journal}{Phys. Rev. Lett.} \textbf{\bibinfo{volume}{107}},
  \bibinfo{pages}{091601} (\bibinfo{year}{2011}).

\bibitem[{\citenamefont{Ding}(2009)}]{Msigma1}
\bibinfo{author}{\bibfnamefont{G.-J.} \bibnamefont{Ding}},
  \bibinfo{journal}{Phys. Rev. D} \textbf{\bibinfo{volume}{79}},
  \bibinfo{pages}{014001} (\bibinfo{year}{2009}).

\end{thebibliography}

\end{document}